\documentclass[runningheads]{llncs}
\usepackage[english]{babel}
\usepackage[T1]{fontenc}
\usepackage[utf8]{inputenc}
\usepackage{graphicx}
\usepackage{subcaption}
\usepackage{caption}
\usepackage{booktabs}
\usepackage{url}
\usepackage{balance}

\usepackage{enumitem}
\newlist{dsitemize}{itemize}{1}
\setlist[dsitemize,1]{label=,leftmargin=0mm}

\usepackage{graphicx}
\usepackage{todonotes}
\usepackage{multirow}
\usepackage{tikz}
\usetikzlibrary{arrows.meta}
\usepackage{amsmath}
\usepackage{amssymb}
\usepackage{amsfonts}
\usepackage{thmtools}		
\usepackage{mleftright}
\usepackage{thm-restate}
\usepackage[mathic=true]{mathtools}
\usepackage{fixmath} 

\usepackage{enumitem}
\usepackage{ellipsis}
\usepackage{xspace}
\usepackage{algorithm}
\usepackage{algorithmicx}
\usepackage{algpseudocode}
\algdef{SE}{ParForAllLoop}{EndParForAllLoop}[1]{\textbf{parallel for}\xspace\(\mbox{#1}\) \textbf{do}}{\textbf{end}}
\usepackage{ifthen}
\newcommand{\CC}[1][]{$\text{C\hspace{-.25ex}}^{_{_{_{++}}}}
    \ifthenelse{\equal{#1}{}}{}{\text{\hspace{-.625ex}#1}}$}

\DeclarePairedDelimiter{\norm}{\lVert}{\rVert}
\newcommand{\tg}{\mathcal{G}}
\newcommand{\tge}{\mathcal{E}}

\newcommand{\sz}{\scriptstyle }

\newcommand{\tgkwedge}{\mbox{\emph{TGK}-$\wedge$}\xspace}
\newcommand{\tgkstar}{\mbox{\emph{TGK}-$\star$}\xspace}

\newcommand{\tgkall}{\mbox{\emph{TGK-all}}\xspace}
\newcommand{\tgkap}{\mbox{\emph{Approx-}}}

\usepackage{hyperref}
\usepackage{cleveref}

\usepackage{tikz}
\usepackage{wrapfig}

\usetikzlibrary{arrows,arrows.meta,decorations.pathmorphing}
\tikzset{
    nodeblack/.style={fill,circle,inner sep=01pt,minimum size=2mm},
    dedge3/.style={->,> = latex',font=\footnotesize,thick},
}

\usepackage{multibib}
\newcites{app}{Appendix References}%
\title{A Temporal Graphlet Kernel for Classifying Dissemination in Evolving Networks}
\titlerunning{A Temporal Graphlet Kernel}
\author{Lutz Oettershagen\inst{1} \and Nils M. Kriege\inst{2,3} \and Claude Jordan\inst{1} \and Petra Mutzel\inst{1}}
\authorrunning{L. Oettershagen et al.}
\institute{Institute of Computer Science, University of Bonn, Bonn, Germany\\
    \email{\{lutz.oettershagen, petra.mutzel\}@cs.uni-bonn.de}, \email{claudejordan@gmail.com} \and
    Faculty of Computer Science, University of Vienna, Vienna, Austria\\\email{nils.kriege@univie.ac.at} \and
    Research Network Data Science, University of Vienna, Vienna, Austria
}

\begin{document}
\maketitle

\begin{abstract}%
We introduce the \emph{temporal graphlet kernel} for classifying dissemination processes in labeled temporal graphs. Such dissemination processes can be spreading (fake) news, infectious diseases, or computer viruses in dynamic networks. The networks are modeled as labeled temporal graphs, in which the edges exist at specific points in time, and node labels change over time. The classification problem asks to discriminate dissemination processes of different origins or parameters, e.g., infectious diseases with different infection probabilities.  
Our new kernel represents labeled temporal graphs in the feature space of temporal graphlets, i.e., small subgraphs distinguished by their structure, time-dependent node labels, and chronological order of edges.
We introduce variants of our kernel based on classes of graphlets that are efficiently countable.
For the case of temporal wedges, we propose a highly efficient approximative kernel with low error in expectation.
We show that our kernels are faster to compute and provide better accuracy than state-of-the-art methods.


\textbf{Kind of paper:} Novel research paper
\end{abstract}

\section{Introduction}
\begin{figure}[t]\centering
    \includegraphics[width=.85\linewidth]{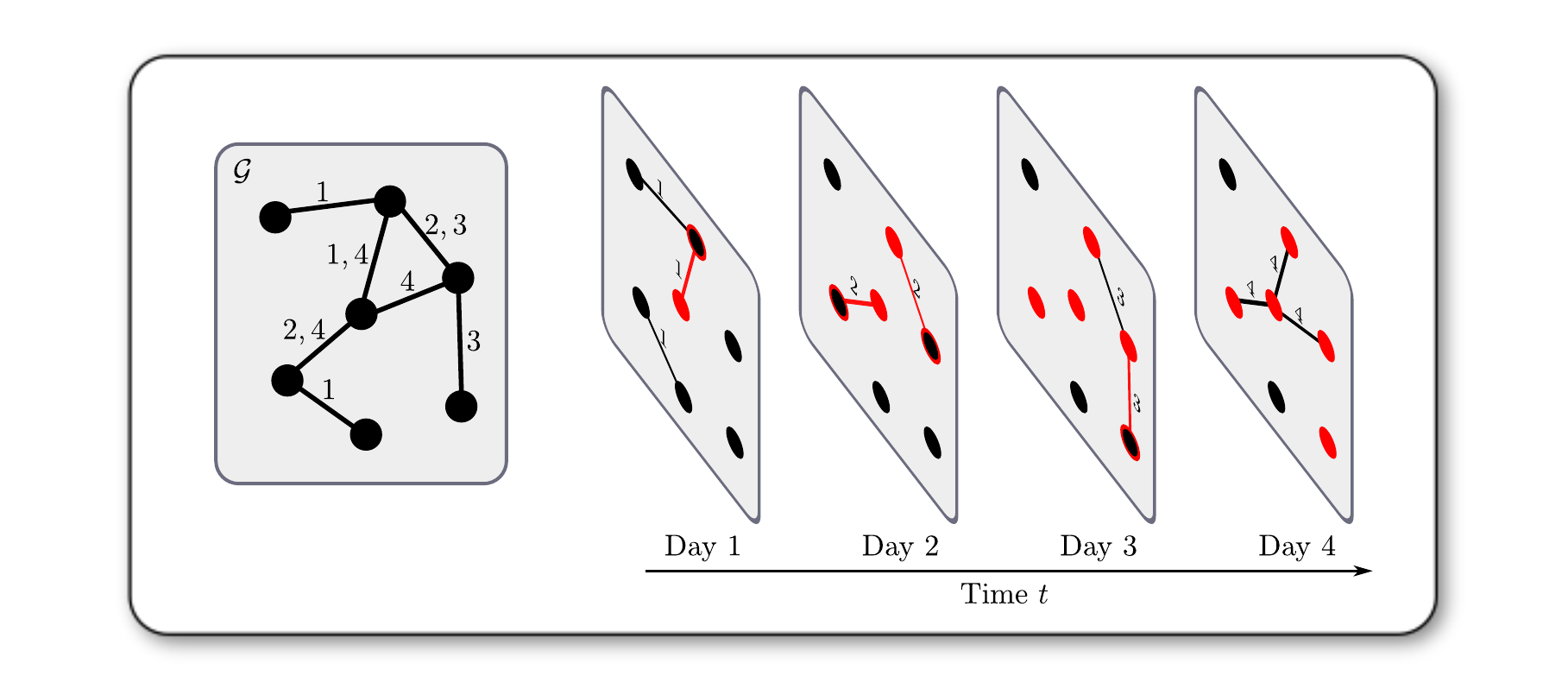}
    \caption{Example for temporal graph $\tg$ in which nodes represent persons and edges contacts over four days. 
        The edge labels denote on which days the edges are available.
        An infection spreads in the network: On the first day, one person is initially infected (highlighted in red). 
        The infection spreads along the edges highlighted in red. The next infected persons are outlined in red.}
    \label{fig:example_spread}
    \vspace{-3mm}
\end{figure}

Dissemination processes such as spreading information or disease can be challenging to analyze and track.
Recent works \cite{oettershagen2020classifying,oettershagen2020temporal,Tortorella2021} discuss the problem of classifying 
dissemination processes in social and human contact networks, e.g., discriminating the spread of real news from fake news, different infectious diseases, or malicious from benign network communications. 
Identifying such spreading processes correctly in real-world and online social networks can have immense social impacts.
For example, detecting a new viral pathogen spreading in communities can help to react early and prevent severe outbreaks~\cite{brouwer2018epidemiology,palladino2020excess}. 
Similarly, identifying and limiting the spread of fake news on social networks like \emph{Facebook} or \emph{WeChat} can help to reduce resulting social unrest~\cite{murayama2021modeling,shu2017fake,vosoughi2018spread}.
These kinds of spreading processes can naturally be modeled using labeled temporal graphs.
A labeled temporal graph consists of a fixed set of nodes, a label function that assigns a discrete label to each node at each point in time, and a set of timestamped edges. 
\Cref{fig:example_spread} shows an example of a temporal graph and a dissemination process over time (highlighted in red).
Because temporal graphs are suitable models for a wide range of real-life scenarios with dynamic relations~\cite{holme2015modern},
research in temporal graphs has recently gained increasing attention~\cite{braha2009time,masuda2019detecting,oettershagen2020classifying,paranjape2017motifs}. 
Methods for analyzing and classifying static graphs, on the other hand, have been studied for decades and are an established area of research---dominating learning methods are based on graph kernels~\cite{Kriege2019} and graph neural networks~\cite{wu2019comprehensive}. 
A well-known and popular static graph kernel is the \emph{graphlet kernel} proposed in~\cite{shervashidze2009efficient}.
We introduce a temporal version, named \emph{temporal graphlet kernel}, to account for the specific properties of temporal graphs. Our kernel is inspired by recent works on temporal motifs~\cite{liu2019sampling,liu2021temporal,mackey2018chronological,paranjape2017motifs,sarpe2021oden}.
Temporal motifs are generalizations of small subgraph patterns, i.e., graphlets, that incorporate  temporal properties like the chronological ordering of the edges. 
In contrast to previous works~\cite{oettershagen2020classifying,oettershagen2020temporal}, our {temporal graphlet kernel} directly operates on the labeled temporal graphs without prior transformations into static graphs. 

The authors of \cite{oettershagen2020temporal} proposed three classification tasks of dissemination on temporal networks.
The first one concerns the discrimination of observations of a dissemination process and random observations.
The second task aims to discriminate temporal graphs subject to two dissemination processes differing in 
the infection probability. Finally, the third classification task is similar to the second one, but under incomplete information, i.e., parts of the network's observations are unavailable.  
Our kernel achieves a high state-of-the-art classification accuracy and efficiency on all three tasks. 

\smallskip
\noindent\textbf{Contributions:}
\begin{itemize}[noitemsep,topsep=0pt]
  \item We introduce the temporal graphlet kernel, defined as the inner product of normalized graph feature vectors counting the occurrences of labeled temporal graphlets. 
  \item We present a highly efficient approximation variant of our temporal graphlet kernel that approximates the number of temporal wedges.
  \item We show that our temporal graphlet kernels reach high accuracies in classifying epidemic spreading in real-world data sets. In most cases, our temporal graphlet kernels beat the state-of-the-art kernels and neural networks and are on par otherwise. The gains in accuracy are often significant. Moreover, our kernels are up to several orders of magnitude faster than the baselines.
\end{itemize}

\medskip
\noindent
We give an overview of the related work in~\Cref{sec:relatedwork}.

\section{Preliminaries}

A \emph{labeled, temporal graph} $\tg=(V, \tge, l)$ consists of a finite set $V$ of nodes, a finite set $\tge$ of (directed) \emph{temporal edges} $e=(u,v,t)$ with $u$ and $v$ in $V$, $u\neq v$, \emph{availability time} (or \emph{time stamp}) $t \in \mathbb{N}$, and a labeling function $l \colon V\times \mathbb{N} \to \Sigma$. The labeling function $l$ assigns a label to each node at each time step $t\in \mathbb{N}$.  
Let $d(v)$ be the degree of vertex $v$, i.e., the total number of incoming and outgoing temporal edges.
We denote with $S(\tg)=(V,E)$ the underlying static graph of the temporal graph $\tg=(V,\tge)$ with $E=\{(u,v)\mid (u,v,t)\in \tge\}$.

\smallskip
\noindent
\textbf{Kernels for Graphs:~~}%
A \emph{kernel} on a non-empty set $\mathcal{X}$ is a symmetric, positive semidefinite function 
$k \colon \mathcal{X} \times \mathcal{X} \to \mathbb{R}$.
Equivalently, a function $k$ is a kernel if there is a \emph{feature map} 
$\phi \colon \mathcal{X} \to \mathcal{H}$ to a Hilbert space $\mathcal{H}$ with inner product 
$\langle \cdot, \cdot \rangle$, such that 
$k(x,y) = \langle \phi(x),\phi(y) \rangle$ for all $x$ and $y$ in $\mathcal{X}$.
Let $\mathbb{G}$ be the set of all (temporal) graphs, then a kernel $k \colon \mathbb{G} \times \mathbb{G} \to \mathbb{R}$ is a \emph{(temporal) graph kernel}.

\smallskip
\noindent
\textbf{Static Graphlet Kernel:~~}
Shervashidze et al. \cite{shervashidze2009efficient} introduced the graphlet kernel for static (unlabeled) graphs.
It counts the occurrences of subgraph patterns of a fixed size which are called \emph{graphlets}.
For a graph $G$ and $k \in \{3,4,5\}$, the static graphlet kernel counts the isomorphism types of all induced 
subgraphs with $k$ nodes. The subgraphs can be disconnected.
Let $\phi(G)_{\sigma_i}$ be the number of occurrences of the isomorphism type $\sigma_i$ for 
$1\leq i \leq N$ with $N$ the number of different types.  
The feature map of the kernel is then $\phi_{SG}(G)=(\phi(G)_{\sigma_1},\ldots, \phi(G)_{\sigma_N})$, and 
the graphlet kernel is $k_{SG}(G,H)=\langle \phi_{SG}(G),\phi_{SG}(H) \rangle$ for all graphs $G$ and $H$ in $\mathbb{G}$.
For small graphs representing molecules, labeled graphlets have also been considered~\cite{wale2008comparison}.

\smallskip
\noindent
\textbf{Temporal Motifs:~~}
Our work is based on extending the commonly used definition of temporal motifs first introduced in \cite{paranjape2017motifs}.

\begin{definition}\label{def:tempgraphlet}
    A \emph{$k$-node, $\ell$-edge, $\delta$-temporal graphlet} is a sequence of $\ell$ temporal edges, 
    $g=\left((u_1 , v_1 , t_1 ), (u_2 , v_2 , t_2 ),\ldots , (u_\ell , v_\ell , t_\ell )\right)$ 
    that is 
    (i) chronologically ordered, i.e., $t_1 < t_2 < \cdots < t_\ell$, 
    (ii) within a $\delta$ time interval, i.e., $t_\ell-t_1\leq \delta$, and 
    (iii) the induced static graph is connected and has $k$ nodes.
\end{definition}

\section{Temporal Graphlet Kernel}

We extend \Cref{def:tempgraphlet} for labeled temporal graphs.
Our approach is general and can be adapted for other definitions.

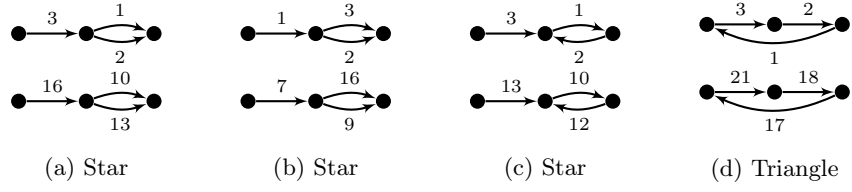
\begin{figure}[t]\centering
    \begin{subfigure}{0.25\columnwidth}\centering
        \begin{tikzpicture}[scale=0.9]
        \node[nodeblack] (1) at (0,1) {};
        \node[nodeblack] (2) at (1,1) {};
        \node[nodeblack] (3) at (2,1) {};
   
        \path[->] 
        (1)  edge[dedge3] node[midway,above] {\scriptsize$3$} (2)
        (2)  edge[dedge3,bend right=20] node[midway,below] {\scriptsize$2$} (3)
        (2)  edge[dedge3,bend left=20] node[midway,above] {\scriptsize$1$} (3);    
        
        \node[nodeblack] (1) at (0,0) {};
        \node[nodeblack] (2) at (1,0) {};
        \node[nodeblack] (3) at (2,0) {};
        
        \path[->] 
        (1)  edge[dedge3] node[midway,above] {\scriptsize$16$} (2)
        (2)  edge[dedge3,bend right=20] node[midway,below] {\scriptsize$13$} (3)
        (2)  edge[dedge3,bend left=20] node[midway,above] {\scriptsize$10$} (3);   
        \end{tikzpicture}
        \subcaption{Star}
        \label{fig:star_graphlet_a}
    \end{subfigure}%
    \begin{subfigure}{0.25\columnwidth}\centering
        \begin{tikzpicture}[scale=0.9]
        \node[nodeblack] (1) at (0,1) {};
        \node[nodeblack] (2) at (1,1) {};
        \node[nodeblack] (3) at (2,1) {};
               
       \path[->] 
       (1)  edge[dedge3] node[midway,above] {\scriptsize$1$} (2)
       (2)  edge[dedge3,bend right=20] node[midway,below] {\scriptsize$2$} (3)
       (2)  edge[dedge3,bend left=20] node[midway,above] {\scriptsize$3$} (3)
       ;      
       
       \node[nodeblack] (1) at (0,0) {};
       \node[nodeblack] (2) at (1,0) {};
       \node[nodeblack] (3) at (2,0) {};
       
       \path[->] 
       (1)  edge[dedge3] node[midway,above] {\scriptsize$7$} (2)
       (2)  edge[dedge3,bend right=20] node[midway,below] {\scriptsize$9$} (3)
       (2)  edge[dedge3,bend left=20] node[midway,above] {\scriptsize$16$} (3)
       ;        
        \end{tikzpicture}
        \subcaption{Star}
        \label{fig:star_graphlet_b}
    \end{subfigure}\hfill%
    \begin{subfigure}{0.25\columnwidth}\centering
        \begin{tikzpicture}[scale=0.9]
        \node[nodeblack] (1) at (0,1) {};
        \node[nodeblack] (2) at (1,1) {};
        \node[nodeblack] (3) at (2,1) {};
        
        \path[->] 
        (1)  edge[dedge3] node[midway,above] {\scriptsize$3$} (2)
        (3)  edge[dedge3,bend left=20] node[midway,below] {\scriptsize$2$} (2)
        (2)  edge[dedge3,bend left=20] node[midway,above] {\scriptsize$1$} (3)
        ;   
        
        \node[nodeblack] (1) at (0,0) {};
        \node[nodeblack] (2) at (1,0) {};
        \node[nodeblack] (3) at (2,0) {};
        
        \path[->] 
        (1)  edge[dedge3] node[midway,above] {\scriptsize$13$} (2)
        (3)  edge[dedge3,bend left=20] node[midway,below] {\scriptsize$12$} (2)
        (2)  edge[dedge3,bend left=20] node[midway,above] {\scriptsize$10$} (3)
        ;       
        
        \end{tikzpicture}
        \subcaption{Star}
        \label{fig:star_graphlet_c}
    \end{subfigure}\hfill%
    \begin{subfigure}{0.25\columnwidth}\centering
        \begin{tikzpicture}[scale=0.9]
        \node[nodeblack] (1) at (0,1) {};
        \node[nodeblack] (2) at (1,1) {};
        \node[nodeblack] (3) at (2,1) {};
        
        \path[->] 
        (1)  edge[dedge3] node[midway,above] {\scriptsize$3$} (2)
        (2)  edge[dedge3] node[midway,above] {\scriptsize$2$} (3)
        (3)  edge[dedge3,bend left=25] node[midway,below] {\scriptsize$1$} (1)
        ;       
        \node[nodeblack] (1) at (0,0) {};
        \node[nodeblack] (2) at (1,0) {};
        \node[nodeblack] (3) at (2,0) {};
        
        \path[->] 
        (1)  edge[dedge3] node[midway,above] {\scriptsize$21$} (2)
        (2)  edge[dedge3] node[midway,above] {\scriptsize$18$} (3)
        (3)  edge[dedge3,bend left=25] node[midway,below] {\scriptsize$17$} (1)
        ;    
        
        \end{tikzpicture}
        \subcaption{Triangle}
        \label{fig:star_graphlet_d}
    \end{subfigure}\hfill
    \caption{Examples of $3$-node, $3$-edge, $10$-temporal graphlets without labels.
    The two graphlets in each of (a)--(d) are equivalent. All other pairs of graphlets are non-equivalent, e.g., the graphlets in (a) are non-equivalent to those in (b)--(d). }        
    \label{fig:graphlets}
\end{figure}

\begin{definition}
   In a labeled temporal graph, the label of a $k$-node, $\ell$-edge, $\delta$-temporal graphlet 
   $g=\left((u_1 , v_1 , t_1 ), (u_2 , v_2 , t_2 ),\ldots , (u_\ell , v_\ell , t_\ell )\right)$ is
   $$l(g)=(l(u_1,t_1), l(v_1,t_1+1), l(u_2,t_2), l(v_2,t_2+1), \ldots, l(u_\ell,t_\ell), l(v_\ell,t_{\ell}+1)).$$
\end{definition}
We are interested in the classification of dissemination that spreads along temporal edges. 
The labels can be used to encode different dissemination patterns. 
Next, we define an equivalence relation on graphlets.
\begin{definition}
    Two $\delta$-temporal graphlets  $g=\left((u_1 , v_1 , t_1 ), \ldots , (u_\ell , v_\ell , t_\ell )\right)$
    and $g'=\left((u'_1 , v'_1 , t'_1 ),\ldots , (u'_{\ell'} , v'_{\ell'} , t'_{\ell'} )\right)$
    are equivalent, written $g\sim g'$, if 
    (i) $\ell=\ell'$, 
    (ii) there exists a bijection $\psi$ from the nodes of $g$ to the nodes of $g'$ with $\psi(u_i)=u'_i$ and $\psi(v_i)=v'_i$ for all $i\in\{1,\ldots,\ell\}$, and,
    in case of labels, (iii) $l(g)=l(g')$.
\end{definition}
A temporal graphlet is edge-induced, in contrast to the static graphlet kernel, which counts the number of node-induced, possibly disconnected, subgraphs.
Moreover, our equivalence relation considers the chronological order of the edges but not the actual time stamps, which would be too restrictive. 
\Cref{fig:graphlets} shows an example for the equivalence relation. Note that non-equivalence can also arise from different labels in the case of labeled temporal graphs. Next, we use the equivalence relation to define graph feature vectors based on temporal graphlets.
\begin{definition}
 For a parameter $\delta$ in $\mathbb{N}$, let $\mathcal{T}$ be the equivalence classes of $\sim$.
 Given a temporal graph $\tg$, we define $\hat{\phi}^\mathcal{T}_\text{TG}(\tg)$ as the vector
 with a component for each $\tau\in\mathcal{T}$ counting the occurrences of temporal graphlets equivalent to $\tau$ in $\tg$. We denote the normalized feature vector by $\phi^\mathcal{T}_\text{TG}(\tg) = \frac{\hat{\phi}^\mathcal{T}_\text{TG}(\tg)}{\norm{\hat{\phi}^\mathcal{T}_\text{TG}(\tg)}_1}$.
\end{definition}

We derive the temporal graph kernel from these feature vector.

\begin{definition}[Temporal Graphlet Kernel]
    Given two temporal graphs $\tg_1$ and $\tg_2$, the \emph{temporal graphlet kernel} is 
    $k^\mathcal{T}_\text{TG}(\tg_1, \tg_2) = \langle \phi^\mathcal{T}_\text{TG}(\tg_1),\phi^\mathcal{T}_\text{TG}(\tg_2) \rangle$.
\end{definition}

\section{Counting Temporal Graphlets}
Counting temporal graphlets is a hard problem.
In general, deciding if a $\delta,k$-star graphlet exists in a given temporal graph is \texttt{NP}-complete~\cite{liu2019sampling}.
The known counting and enumeration algorithms for general temporal graphlets have exponential worst-case running times, e.g., the backtracking enumeration algorithm
of Mackey et al.~\cite{mackey2018chronological} or the general counting algorithm presented in~\cite{paranjape2017motifs}.
Besides these negative results, several special cases of temporal graphlets 
can be computed efficiently.
In the following, we discuss the counting of general temporal graphlets and the
efficient counting of temporal graphlets with two or three nodes and edges, respectively, which are elementary classes of motifs for 
the characterization of temporal networks~\cite{uvzupyte2020test,paranjape2017motifs}, as well as the underlying dissemination process.

\subsection{Labeled Temporal Graphlets for Dissemination}
\begin{figure}[t]\centering
    \begin{subfigure}{0.24\columnwidth}\centering
        \begin{tikzpicture}[scale=0.9]
        \node[nodeblack] (1) at (0,1) {};
        \node[nodeblack] (2) at (1,1) {};
        \node[nodeblack] (3) at (2,1) {};
        
        \path[->] 
        (1)  edge[dedge3] node[midway,above] {\scriptsize$t_1$} (2)
        (2)  edge[dedge3] node[midway,above] {\scriptsize$t_2$} (3);

        \end{tikzpicture}
    \end{subfigure}%
    \begin{subfigure}{0.24\columnwidth}\centering
        \begin{tikzpicture}[scale=0.9]
        
        \node[nodeblack] (1) at (0,0) {};
        \node[nodeblack] (2) at (1,0) {};
        \node[nodeblack] (3) at (2,0) {};
        
        \path[->] 
        (2)  edge[dedge3] node[midway,above] {\scriptsize$t_1$} (1)
        (3)  edge[dedge3] node[midway,above] {\scriptsize$t_2$} (2);   
        \end{tikzpicture}
    \end{subfigure}%
    \begin{subfigure}{0.24\columnwidth}\centering
        \begin{tikzpicture}[scale=0.9]
        \node[nodeblack] (1) at (0,1) {};
        \node[nodeblack] (2) at (1,1) {};
        \node[nodeblack] (3) at (2,1) {};
        
        \path[->] 
        (1)  edge[dedge3] node[midway,above] {\scriptsize$t_1$} (2)
        (3)  edge[dedge3] node[midway,above] {\scriptsize$t_2$} (2)
        ;      

        \end{tikzpicture}
    \end{subfigure}%
    \begin{subfigure}{0.24\columnwidth}\centering
        \begin{tikzpicture}[scale=0.9]
        \node[nodeblack] (1) at (0,1) {};
        \node[nodeblack] (2) at (1,1) {};
        \node[nodeblack] (3) at (2,1) {};
        
        \path[->] 
        (2)  edge[dedge3] node[midway,above] {\scriptsize$t_1$} (1)
        (2)  edge[dedge3] node[midway,above] {\scriptsize$t_2$} (3)
        ;   
        \end{tikzpicture}
    \end{subfigure}\hfill%
   
    \caption{The four non-equivalent temporal wedges with time stamps $t_1<t_2$.}
    \label{fig:wedges}
\end{figure}
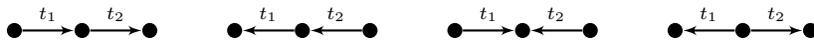

Our goal is to classify disease or information spreading in temporal graphs.  Therefore, we use a binary label alphabet $\Sigma$ that encodes if a node is \emph{infected} or \emph{susceptible} or if a node has obtained some information or not. 
It is possible to use our approach for larger alphabet sizes that include, e.g., \emph{exposed} or \emph{recovered} labels to model complex epidemiological behaviors~\cite{brauer2008compartmental}.
In the following, we discuss the counting of small graphlets with two or three nodes and edges, respectively.
The motivation is that the dissemination of, e.g., viruses or (fake) news on social networks is usually a mainly local  process~\cite{kaslow2014viral,vosoughi2018spread}. Therefore, we expect small and connected graphlets to capture these processes well---we verify this hypothesis in \Cref{sec:experiments}. 
The authors of \cite{paranjape2017motifs} identified 36 non-equivalent temporal unlabeled graphlets for $k\in\{2,3\}$ nodes and $\ell=3$ edges. More specifically, eight non-equivalent triangles and 24 stars with three edges.
In the case of three vertices and two edges, the temporal graphlet is called a \emph{temporal wedge}. \Cref{fig:wedges} shows the four non-equivalent (unlabeled) temporal wedges.
If we take node labelings into account, the number of non-equivalent graphlets increases.
\begin{lemma}\label{theorem:number_of_label_seqs}
    Let $\tg$ be a labeled temporal graph and $L=|\Sigma|$ the size of its label alphabet. 
    The number of distinct labels of $\ell$-edge temporal graphlets is $L^{2\ell}$.
\end{lemma}

It follows from \Cref{theorem:number_of_label_seqs}  that for $\Sigma=\{\text{infectected}, \text{susceptible}\}$ there are a total number of $2304$ pair-wise non-equivalent labeled graphlets in $\mathcal{G}$ with two or three nodes and three edges. Similarly, the number of non-equivalent labeled temporal wedges is $64$.

\subsection{Counting General Temporal Graphlets}
In order to count the occurrences of general labeled graphlets, we adapt the counting algorithm introduced in~\cite{paranjape2017motifs}.
Counting the occurrences of a (non-labeled) temporal graphlet $g$ is done in three steps:
\begin{enumerate}
    \item Find all instances of the subgraph $S(g)$ in $S(\tg)$, where $S(g)$ denotes the underlying static graph of the graphlet $g$, i.e., all temporal edges are replaced by static edges, and multi-edges are merged.
    \item For each $S(g)$, collect all temporal edges incident only to nodes in $S(g)$ in a chronologically ordered sequence $\sigma$. 
    \item Count the occurrences of $g$ in $\sigma$ using dynamic programming.
\end{enumerate}
The first step is done using an algorithm for enumerating subgraphs in the static graph $S(\tg)$. The second step is just collecting the corresponding temporal edges.
The third step uses a dynamic programming approach in which a temporal graphlet is considered as a chronologically ordered sequence of temporal edges.
Therefore, the count of a graphlet of length $\ell$ edges can be computed using the count of the prefix of length $\ell-1$.
To this end, for each possible edge sequence of lengths smaller or equal to $\ell$, a counter keeps track of its occurrences.
Let $m$ be the number of static edges in $S(g)$, then all occurrences of $g$ in the sequence $\sigma$ can be counted with a single pass over the edge sequence $\sigma$ in $\mathcal{O}(m^\ell \cdot |\sigma|)$ running time because each edge leads to maximal $\mathcal{O}(m^\ell)$ counter updates when it is processed and when it leaves the time window $\delta$. The sliding time window is used to ensure that only graphlets are counted that respect the maximal temporal distance of $\delta$.
Furthermore, after $\sigma$ is processed, the counters contain all occurrences of graphlets with maximal $\ell$ edges in $S(g)$.

In order to apply the counting framework for labeled temporal graphlets, we map the time-dependent node labels to edge labels. 

\begin{lemma}\label{lemma:mapping}
    Given a temporal graphlet $g=(e_1,\ldots,e_\ell)$, its label $l(g)$
    can be mapped one-to-one to an edge label sequence $l_e(g)=\left(l_e(e_1),\ldots,l_e(e_\ell)\right)$. 
\end{lemma}
To count the labeled graphlets, we now can apply \Cref{lemma:mapping} to the counting framework, where we  
use, for each possible \emph{labeled edge sequence} $\sigma_l$ of lengths smaller or equal to $\ell$, a counter that keeps track of the occurrences of $\sigma_l$.
The running time for counting labeled temporal graphlets is then in $\mathcal{O}((m\cdot |\Sigma|^2)^\ell\cdot |\sigma_l|)$.

\subsection{Counting Two and Three Node Graphlets}
The running time of the general counting algorithm described in the previous section can be improved for graphlets with two or three nodes and three edges. 
Efficient variants for graphlets with two or three nodes and three edges exist, namely for \emph{stars} and \emph{triangles}~\cite{paranjape2017motifs}.
Counting the number of all labeled star graphlets over three nodes and edges is possible in $\mathcal{O}(|\tge|\cdot |\Sigma|^2)$.
We can count temporal wedges similarly. 
And, counting the numbers of all labeled temporal triangles has a running time in 
$\mathcal{O}(\text{T}_\Delta + c_\Delta|\tge|\cdot|\Sigma|^2)$, where $\text{T}_\Delta$ is the running time for the enumeration of static triangles in $S(\tg)$ and $c_\Delta$ the number of static triangles in $S(\tg)$.
We refer the reader to \cite{paranjape2017motifs} for further details of the algorithms.

\subsection{Temporal Wedge Kernel Approximation}
As we will see in \Cref{sec:experiments}, the wedge-based temporal graphlet kernel shows an excellent trade-off between running time and classification accuracy. 
The reason is that the temporal wedges can capture dissemination patterns well, and they are easy to count.
We further improve this trade-off with a highly efficient approximation algorithm.
First, note that we can count the labeled wedges by iterating over all nodes and considering all incoming and outgoing edge combinations.
For each pair of edges, we increase the counter of the correspondent labeled wedge. 
The running time for counting all labeled wedges is then in 
$\mathcal{O}(\sum_{v\in V}{d(v)^2})$.

We introduce an approximation for the temporal wedge kernel based on this counting variant.  
The idea is to sample temporal wedges by extending techniques for wedge sampling in static graphs~\cite{seshadhri2014wedge}.
First, let $\tg=(V,\tge, l)$ be a labeled temporal graph that has no vertex with two incident edges with the same time stamp (we discuss the general case later). Then $w=\sum_{v\in V}{d(v) \choose 2}$ is the total number of temporal wedges in $\tg$. 
We first sample a vertex with a probability of $p_v={d(v)\choose 2}/w$ and then a pair of to $v$ incident edges uniformly at random, i.e., with probability $1/{d(v)\choose 2}$. The probability of the sampled temporal  wedge is then $1/w$, and therefore, it is a uniform sample.
\Cref{alg:as} shows the approximation for the feature vector $\widetilde{\phi}_{\text{TG}}(\tg)$ of normalized temporal wedge counts. 

\begin{theorem}\label{theorem:runtime}
    The running time of \Cref{alg:as} is in $\mathcal{O}(|V|+|\Sigma|^{2}+s)$.
\end{theorem}

\vspace{-5mm}
\begin{algorithm}\mbox{\hfill}
    \\\textbf{Input:} A temporal graph $\tg$, sample size $s\in\mathbb{N}$ %
    \\\textbf{Output:} A feature vector $\widetilde{\phi}_{\text{TG}}(\tg)$ of normalized temporal wedge counts
    \begin{algorithmic}[1]
        \State Initialize feature vector $\widetilde{\phi}_{\text{TG}}(\tg)$ to a vector of zeros
        \State Compute probabilities $p_v$ for all $v\in V$
        \For{$1,\ldots,s$}
        \State Sample vertex $v$ with probability $p_v$
        \State Sample two edges $\{e,f\}$ incident to $v$ %
        \State Let $\tau$ be the temporal wedge defined by $e$ and $f$
        \State $\widetilde{\phi}_{\text{TG}}(\tg)_{\tau} \gets \widetilde{\phi}_{\text{TG}}(\tg)_{\tau} + \frac{1}{s}$
        \EndFor
        \State \Return$\widetilde{\phi}_{\text{TG}}(\tg)$ %
    \end{algorithmic}
    \caption{}
    \label{alg:as}
\end{algorithm}
\vspace{-5mm}

\begin{theorem}\label{tgclass:theorem:approx}
    Let $\mathbb{G}$ be a set of temporal graphs with label alphabet $\Sigma$. 
    Moreover, let $W$ be number of non-equivalent labeled temporal wedges, $\lambda\in\mathbb{R}_{>0}$, and $\delta\in(0,1)$. 
    For 
    $s = \left\lceil\frac{\log( 2 \cdot  |\mathbb{G}| \cdot W/\delta)}{ 2 ({\lambda}/{W})^2  }\right\rceil$,
    \Cref{alg:as} approximates the normalized temporal graphlet wedge kernel ${k}_{\text{TG}}$ with probability $(1-\delta)$, such that 
    \begin{equation*}
    \sup_{\tg_1, \tg_2 \in \mathbb{H}} \left|    {k}_{\text{TG}}(\tg_1, \tg_2) - \langle \widetilde{\phi}_{\text{TG}}(\tg_1), \widetilde{\phi}_{\text{TG}}(\tg_2)  \rangle  \right| \leq 3\lambda.
    \end{equation*}
\end{theorem}
To support temporal graphs that have equal time stamps of edges at the same vertex, and to obtain a random uniform temporal wedge respecting a given time interval $\delta$, 
we sample a vertex $v$ with probability $p_v$ and one of the edges incident to $v$ uniformly at random.
Then we choose the second edge, $f$, only from the incident edges such that the difference of the availability times of $e$ and $f$ is in $[1,\delta]$.
To correct the introduced bias, we apply rejection sampling.
Let $p_f$ be the probability of the second edge $f$.
The probability of a temporal wedge $\tau$  respecting $\delta$ and with center $v$ is then $P_\tau=\frac{p_v\cdot  p_f}{d(v)}$. 
A lower bound for the probability of any temporal wedge is $P_{min}={1}/{w}$.
To obtain a uniformly sampled temporal wedge respecting the interval $\delta$, we accept a wedge with probability $\frac{P_{min}}{P_\tau}$, and we achieve a uniform probability of $P_\tau \cdot \frac{P_{min}}{P_{\tau}}=1/w$.

If the input graphs have similar distributions of the edge times and degrees, forgoing the rejection step leads to similar biases during the wedge sampling. In this case, we may waive the rejection step and still achieve high accuracy.

\section{Experiments}\label{sec:experiments}
We evaluate our temporal graphlet kernel and compare the effectiveness and efficiency to the baselines provided in \cite{oettershagen2020classifying,oettershagen2020temporal,Tortorella2021}.
Our research questions are:
\begin{itemize}
    \item \textbf{Q1.} How do our new kernels compare to the baselines in terms of accuracy? 
    \item \textbf{Q2.} What are the running times of our temporal graphlet kernels? 
    \item \textbf{Q3.} What is the solution quality and running time of our approximation? 
    \item \textbf{Q4.} How is the classification accuracy affected by incomplete knowledge of the dissemination process? %
\end{itemize}

\noindent\textbf{Data Sets: }
We use the temporal dissemination classification data sets introduced in \cite{oettershagen2020temporal}. The data sets contain the three different classification tasks described in the introduction based on real-world temporal graphs that model physical human interactions and social networks. 
Please refer to \Cref{appendix:datasets} for a detailed description.

\smallskip
\noindent\textbf{Kernel Instances: }
We compare the following variants of our temporal graphlet kernel for $k=3$ nodes and $\ell\in\{2,3\}$ edges: 
1) \emph{TGK}-$\wedge$ counts labeled temporal wedges, 2) \emph{TGK}-$\star$ counts labeled temporal star graphlets, 3) \emph{TGK-all} counts all labeled temporal graphlets, and 4) \emph{Approx-$s$} is our approximation of \emph{TGK}-$\wedge$ with sample size $s\in\{50,100,200\}$.
Note that setting $k=\{2,3\}$, i.e., the kernels also count graphlets on two nodes, did not improve the results.

Furthermore, we use state-of-the-art approaches based on kernels and neural networks as baselines.
The authors of \cite{oettershagen2020temporal} introduce three different transformations of temporal graphs to static graphs, which differ in the size of the resulting graphs and the amount of loss of temporal information.
After the transformations, they apply static kernels to the static graphs for classifying the dissemination process.
They use the $k$-step random walk and the Weisfeiler-Leman subtree kernel.
This approach results in the following kernel instances: (1) \emph{RD-RW} and \emph{RD-WL}, which use the \emph{reduced graph representation}, (2) \emph{DL-RW} and \emph{DL-WL}, which use the \emph{directed line graph expansion}, and (3)  \emph{SE-RW} and \emph{SE-WL}, which use the \emph{static expansion}.
In \cite{oettershagen2020classifying}, the authors of \cite{oettershagen2020temporal} introduced graph neural networks (GNNs) based on the
three graph transformations. However, the evaluation in \cite{oettershagen2020classifying} showed that the kernel-based approaches outperform the GNNs in almost all instances while requiring less computation time. Tortorella and Micheli~\cite{Tortorella2021} introduced a dynamic echo state network called \emph{DynGESN} for classification of dissemination in temporal networks.
Further details and a comparison of our kernels with these neural baselines can be found in \Cref{appendix:gnn}.

We implemented our kernels in C++ using GNU CC Compiler 9.3.0. 
The source code and data sets are available online.\footnote{\url{https://gitlab.com/tgpublic/tgraphlet}}
The C++ implementation of the baseline kernels from \cite{oettershagen2020temporal} were provided by the authors and compiled with the same settings as our kernels.
The experimental protocol is in \Cref{ep}.

\begin{table}[t]\centering
    \caption{Classification accuracy in percent and standard deviation for the first and second classification tasks. For each data set, we highlight the highest accuracy in bold. OOT---Computation did not finish within the time limit.}
    \begin{subtable}{1\textwidth}
    \caption{Classification accuracy for the first classification task. } %
    \label{table:results}
    \resizebox{1\textwidth}{!}{ 	\renewcommand{\arraystretch}{0.9}\setlength{\tabcolsep}{5pt}
        \begin{tabular}{cl@{\hspace{5mm}}ccccccc}	\toprule
            & \multirow{3}{*}{\vspace*{4pt}\textbf{Kernel}}&\multicolumn{6}{c}{\textbf{Data set}}\\\cmidrule{3-8}
            &  & {\textsc{Mit}}         &  {\textsc{Highschool}}      &     {\textsc{Infectious}}       & {\textsc{Tumblr}}       & {\textsc{Dblp}} &  {\textsc{Facebook}} \\ \toprule
            & \tgkwedge    & $\textbf{93.45}$ $\sz\pm 1.8$ & $98.00$ $\sz\pm 0.9$ & $\textbf{98.30}$ $\sz\pm 0.4$ & $93.19$ $\sz\pm 0.7$ & $98.47$ $\sz\pm 0.1$ &  $95.21$ $\sz\pm 0.2$\\
            & \tgkstar     & $87.02$ $\sz\pm 1.6$ & $96.50$ $\sz\pm 0.9$ & ${98.05}$ $\sz\pm 0.4$ & $93.19$ $\sz\pm 0.8$ & $98.05$ $\sz\pm 0.2$ & $95.32$ $\sz\pm 0.2$\\
            & \tgkall      & $87.46$ $\sz\pm 0.9$ & $97.17$ $\sz\pm 0.8$ & $97.90$ $\sz\pm 0.4$ & $93.57$ $\sz\pm 0.8$ & $98.12$ $\sz\pm 0.1$ & $94.69$ $\sz\pm 0.3$\\ \midrule
            & \tgkap$50$   & $88.96$ $\sz\pm 2.1$ & $94.50$ $\sz\pm 0.8$ & $91.75$ $\sz\pm 1.0$ & $91.63$ $\sz\pm 0.4$ & $97.12$ $\sz\pm 0.2$ & $94.05$ $\sz\pm 0.2$\\ 
            & \tgkap$100$  & $89.19$ $\sz\pm 1.6$ & $95.88$ $\sz\pm 1.3$ & $95.15$ $\sz\pm 0.7$ & $91.84$ $\sz\pm 0.6$ & $97.66$ $\sz\pm 0.3$ & $94.10$ $\sz\pm 0.1$\\ 
            & \tgkap$200$  & $90.78$ $\sz\pm 1.5$ & $97.66$ $\sz\pm 0.6$ & $96.80$ $\sz\pm 1.0$ & $92.86$ $\sz\pm 0.5$ & $98.06$ $\sz\pm 0.5$ & $95.02$ $\sz\pm 0.1$\\ 
            \midrule\multirow{6}{*}{\rotatebox{90}{\small Baselines}}
            & \emph{RG-RW} & $61.31$ $\sz\pm 2.7~$ & $90.16$ $\sz\pm 1.0~$ & $89.30$ $\sz\pm 1.0~$ & $74.99$ $\sz\pm 1.9~$ & $90.60$ $\sz\pm 1.0~$ & $82.86$ $\sz\pm 0.6~$ \\
            & \emph{RG-WL} & $81.88$ $\sz\pm 1.1~$ & $89.88$ $\sz\pm 0.9~$ & $91.75$ $\sz\pm 1.0~$ & $70.50$ $\sz\pm 1.0~$ & $90.45$ $\sz\pm 0.5~$ & $81.15$ $\sz\pm 0.8~$ \\
            & \emph{DL-RW} & $92.91$ $\sz\pm 0.9~$ & $98.33$ $\sz\pm 0.7~$ & $97.05$ $\sz\pm 0.8~$ & $\textbf{94.64}$ $\sz\pm 0.5~$ & $98.16$ $\sz\pm 0.1~$ & $96.46$ $\sz\pm 0.1~$ \\
            & \emph{DL-WL} & $90.67$ $\sz\pm 1.6~$ & $\textbf{98.88}$ $\sz\pm 0.4~$ & $97.35$ $\sz\pm 1.5~$ & $94.05$ $\sz\pm 0.9~$ & $98.56$ $\sz\pm 0.3~$ & $\textbf{96.59}$ $\sz\pm 0.4~$ \\
            & \emph{SE-RW} & $88.56$ $\sz\pm 1.0~$ & $96.89$ $\sz\pm 1.2~$ & ${97.60}$ $\sz\pm 0.6~$ & $93.97$ $\sz\pm 0.9~$ & $\textbf{98.65}$ $\sz\pm 0.3~$ & $95.46$ $\sz\pm 0.2~$ \\
            & \emph{SE-WL} & $87.31$ $\sz\pm 1.9~$ & $96.72$ $\sz\pm 0.7~$ & $94.45$ $\sz\pm 1.1~$ & $93.51$ $\sz\pm 0.6~$ & $97.38$ $\sz\pm 0.2~$ & $95.39$ $\sz\pm 0.4~$ \\
            \bottomrule  \end{tabular}}
\end{subtable}
\begin{subtable}{1\textwidth}
    \caption{Classification accuracy for the second classification task.  
    } %
    \label{table:results2}
    \resizebox{1\textwidth}{!}{ 	\renewcommand{\arraystretch}{0.9}\setlength{\tabcolsep}{5pt}
        \begin{tabular}{clccccccc}	\toprule
            & \multirow{3}{*}{\vspace*{4pt}\textbf{Kernel}}&\multicolumn{6}{c}{\textbf{Data set}}\\\cmidrule{3-8}
            &  & {\textsc{Mit}}         &  {\textsc{Highschool}}      &     {\textsc{Infectious}}       & {\textsc{Tumblr}}       & {\textsc{Dblp}} &  {\textsc{Facebook}} \\ \toprule
            & \tgkwedge    & $68.52$ $\sz\pm 3.5$ & $93.83$ $\sz\pm 0.8$ & $89.65$ $\sz\pm 0.8$ & $\textbf{79.06}$ $\sz\pm 0.7$ & $83.76$ $\sz\pm 0.5$ & $76.63$ $\sz\pm 0.3$\\
            & \tgkstar     & $77.03$ $\sz\pm 3.7$ & $\textbf{95.33}$ $\sz\pm 0.8$ & $\textbf{90.55}$ $\sz\pm 1.6$ & ${78.13}$ $\sz\pm 1.1$ & $85.42$ $\sz\pm 0.8$ & $\textbf{82.12}$ $\sz\pm 0.3$ \\
            & \tgkall      & $\textbf{78.43}$ $\sz\pm 4.2$ & $94.22$ $\sz\pm 1.3$ & $90.40$ $\sz\pm 0.9$ & $76.57$ $\sz\pm 1.0$ & $\textbf{85.84}$ $\sz\pm 0.6$ & $81.13$ $\sz\pm 0.6$ \\\midrule
            & \tgkap$50$   & $55.78$ $\sz\pm 3.5$ & $84.77$ $\sz\pm 1.5$ & $84.50$ $\sz\pm 1.1$ & $75.92$ $\sz\pm 0.5$ & $78.00$ $\sz\pm 0,3$ & $74.12$ $\sz\pm 0.4$\\ 
            & \tgkap$100$  & $60.11$ $\sz\pm 4.5$ & $89.61$ $\sz\pm 1.9$ & $85.10$ $\sz\pm 1.0$ & $76.16$ $\sz\pm 0.7$ & $79.32$ $\sz\pm 0.5$ & $74.41$ $\sz\pm 0.3$\\ 
            & \tgkap$200$  & $62.97$ $\sz\pm 2.8$ & $91.94$ $\sz\pm 1.2$ & $85.45$ $\sz\pm 1.4$ & $78.46$ $\sz\pm 0.8$ & $79.61$ $\sz\pm 0.5$ & $76.32$ $\sz\pm 0.3$\\ 
            \midrule
            \multirow{6}{*}{\rotatebox{90}{\small Baselines}}
            & \emph{RG-RW} & $58.03$ $\sz\pm 3.7~$ & $77.33$ $\sz\pm 2.4~$ & $72.05$ $\sz\pm 2.2~$ & $68.48$ $\sz\pm 1.5~$ & $63.24$ $\sz\pm 1.2~$ & $66.68$ $\sz\pm 0.9~$ \\
            & \emph{RG-WL} & $66.81$ $\sz\pm 2.0~$ & $82.78$ $\sz\pm 1.3~$ & $77.40$ $\sz\pm 1.2~$ & $68.25$ $\sz\pm 1.2~$ & $66.16$ $\sz\pm 0.5~$ & $66.96$ $\sz\pm 0.7~$ \\
            & \emph{DL-RW} & OOT & $91.44$ $\sz\pm 1.1~$ & ${87.35}$ $\sz\pm 1.3~$ & $76.51$ $\sz\pm 0.5~$ & $81.79$ $\sz\pm 0.9~$ & ${79.97}$ $\sz\pm 0.5~$ \\
            & \emph{DL-WL} & $40.87$ $\sz\pm 3.6~$ & $87.11$ $\sz\pm 1.7~$ & $77.55$ $\sz\pm 2.0~$ & $78.69$ $\sz\pm 0.8~$ & $74.47$ $\sz\pm 1.1~$ & $79.44$ $\sz\pm 0.5~$ \\
            & \emph{SE-RW} & $51.03$ $\sz\pm 5.1~$ & $90.77$ $\sz\pm 1.1~$ & $83.60$ $\sz\pm 1.1~$ & $77.09$ $\sz\pm 1.0~$ & ${83.31}$ $\sz\pm 1.0~$ & ${79.56}$ $\sz\pm 0.6~$ \\
            & \emph{SE-WL} & $46.52$ $\sz\pm 3.9~$ & ${91.55}$ $\sz\pm 0.9~$ & $79.60$ $\sz\pm 1.5~$ & $78.64$ $\sz\pm 1.4~$ & $81.24$ $\sz\pm 0.6~$ & $74.68$ $\sz\pm 0.7$ \\
            \bottomrule  \end{tabular}}
\end{subtable}
\end{table}

\subsection{Results}

\noindent\textbf{Q1.}
\Cref{table:results} and \Cref{table:results2} show the accuracies for the first and second classification tasks, respectively.
Our temporal graph kernels perform better than the baselines for eight of the twelve data sets and are on par for the remaining four.
For the first classification task, our kernels have the highest accuracies for \textsc{Infectious}, and \tgkwedge also has the highest accuracy for \textsc{Mit}.
For the other data sets, the accuracies are similar to the best performing baselines.
In the case of the second classification task, \tgkwedge, \tgkstar, and \tgkall achieve higher accuracies the baselines for all data sets.
Similarly, \Cref{appendix:gnn} shows that our new kernels usually reach higher accuracies compared to the GNN baselines.
In general, the accuracies of our kernels are very close to each other for most data sets, and they can all capture the discriminating information well.%

\noindent\textbf{Q2.}
\Cref{table:runningtimes2} shows the running times for the kernel computations for the second classification task.
We observed similar results for the first task (see~\Cref{appendix:runningtimes}).
The impact of the time window parameter $\delta$ on the running time of the temporal graphlet kernels is limited.
\tgkwedge is the fastest kernel for all data sets but the \textsc{Mit} data set. Here the WL-kernel based on the reduced graph representation (\emph{RG-WL}) is faster. Its running time also comes close to \tgkwedge for the other data sets, but it has much worse accuracy (see~\textbf{Q1}).
The reason is that there are no multiple edges between pairs of nodes in graphs in reduced graph representation,
which leads to loss of temporal information and affects the accuracy. 
Our \tgkstar is faster than the random walk kernels and the WL-kernels
for \textsc{Highschool} and \textsc{Mit}.
Notice that the running time of \emph{DL-RW} exceeded the time limit of one hour, where our \tgkstar kernel only needed 102 ms. 
\begin{table}[t]\centering	
    \caption{Running times in ms for the second classification task.} 
    \centering
    \begin{subtable}{\linewidth}
        \centering
        \caption{Random walk length $k=3$ ($k=2$ for \emph{DL-RW}), number of iterations of WL $h=3$ ($h=2$ for \emph{DL-WL}). OOT---Computation did not finish within the time limit.} 
        \label{table:runningtimes2}
        \resizebox{0.88\textwidth}{!}{ 	\renewcommand{\arraystretch}{1}\setlength{\tabcolsep}{5pt}
            \begin{tabular}{cc@{\hspace{5mm}}l@{\hspace{5mm}}ccccccc@{}}	\toprule
                
                & \multirow{3}{*}{\vspace*{4pt}\textbf{Kernel}}&&\multicolumn{6}{c}{\textbf{Data set}}\\\cmidrule{4-9}
                & &  & {\textsc{Mit}}        & \textsc{Highschool}  & \textsc{Infectious}&  \textsc{Tumblr}   & \textsc{Dblp}  & \textsc{Facebook}  \\ \toprule
                & \tgkwedge &    $\delta=10$          &  $290$ & $\textbf{36}$  & $\textbf{39}$     & $\textbf{69}$  & $293$ & $\textbf{459}$ \\
                & \tgkstar & $\delta=10$   &  $102$ & $192$ & $202$    & $467$ & $1\,605$ & $2\,819$ \\
                & \tgkall  &     $\delta=10$          &  $311$ & $755$ & $1\,442$ & $815$ & $5\,154$ & $3\,935$ \\
                \midrule
                & \tgkwedge &   $\delta=10^3$           &  $482$ & $56$  & $41$     & $70$  & $\textbf{287}$ & ${463}$ \\
                & \tgkstar & $\delta=10^3$ &  $103$ & $200$ & $211$    & $478$ & $1\,726$ & $2\,796$ \\
                & \tgkall  &   $\delta=10^3$            &  $283$ & $742$ & $1\,608$ & $799$ & $4\,809$ & $4\,024$ \\\cmidrule{1-9}
                \multirow{6}{*}{\rotatebox{90}{\small Baselines}}       
                & \emph{RG-RW} & & $803$     & $9\,868$  & $10\,504$ & $5\,653$   & $17\,516$  & $10\,444$ \\
                & \emph{RG-WL} & & $\bf{14}$ & $62$ & $76$ & $109$ & $304$ & $580$ \\
                & \emph{DL-RW} & & OOT       & $17\,879$ & $5\,747$  & $1\,555$   & $3\,523$   & $2\,344$ \\
                & \emph{DL-WL} & & $56\,877$ & $1\,331$  & $577$     & $312$      & $906$      & $812$ \\
                & \emph{SE-RW} & & $3\,901$  & $1\,887$  & $2\,464$  & $984$      & $3\,359$   & $2\,858$ \\
                & \emph{SE-WL} & & $339$     & $268$     & $193$     & $218$      & $673$      & $820$ \\
                \bottomrule  \end{tabular}}   
            \vspace{-3mm} 
    \end{subtable}
    \begin{subtable}{\linewidth}
        \centering
        \caption{Running times in ms of \emph{Approx}-$s$.}
        \label{table:runningtimesapprox}
        \resizebox{0.8\textwidth}{!}{ 	\renewcommand{\arraystretch}{1}\setlength{\tabcolsep}{5pt}
            \begin{tabular}{l@{\hspace{5mm}}cccccc}	\toprule
                
                \multirow{3}{*}{\vspace*{4pt}\textbf{Kernel}}&\multicolumn{6}{c}{\textbf{Data set}}\\\cmidrule{2-7}
                & {\textsc{Mit}}     &    \textsc{Highschool}  & \textsc{Infectious}&  \textsc{Tumblr}   & \textsc{Dblp}  & \textsc{Facebook}  \\ \toprule
                \emph{Approx}-$50$       &  $5$ & $16$ & $19$ & $66$ & $264$ & $443$ \\
                \emph{Approx}-$100$      &  $6$ & $17$ & $20$ & $67$ & $267$ & $458$ \\
                \emph{Approx}-$200$      &  $7$ & $19$ & $22$ & $71$ & $268$ & $466$ \\
                \bottomrule  \end{tabular}}
    \end{subtable}
    \vspace{-3mm}
\end{table}
\begin{table}[t]
    \centering
    \caption{Running times and accuracy in percent for the synthetic large data set.}
    \label{table:rnd}
    \resizebox{1\textwidth}{!}{ 	\renewcommand{\arraystretch}{1}\setlength{\tabcolsep}{5pt}
        \begin{tabular}{l@{\hspace{5mm}}ccccccc}	\toprule

            &\emph{NTWL} & \emph{SEWL}  & \tgkstar  &  \tgkwedge  & \tgkap100   & \tgkap1000  & \tgkap10000  \\ \toprule
            Run. time   & 10.21 s & 36.34 s & 19.02 s & 3.6 s & 13 ms & 42 ms & 240 ms \\
            Acc.       & $36.32\sz\pm 2.7$ & $90.31\sz\pm 1.3$& $91.61\sz\pm 0.6$ & $91.73\sz\pm 0.6$ & $87.91\sz\pm 1.4$  & $91.63 \sz\pm 0.5$ & $91.73 \sz\pm 0.4$ \\
            \bottomrule  \end{tabular}}
\end{table}

\noindent\textbf{Q3.}
\Cref{table:results} and \Cref{table:results2} show the accuracies of our approximation algorithm for sample sizes $s\in\{50,100,200\}$.
With increasing sample size, the approximation error is reduced as expected.
The approximation error is generally low, while the speed-up compared to \tgkwedge for low sample sizes is high.
However, for $s=200$, the difference in running time of the approximation and \tgkwedge is small for \textsc{Tumblr} and \textsc{Dblp}. 
The reason is that the exact algorithm is already very efficient due to the low average maximal degree in these data sets (see \Cref{table:sizes1}).
Due to \Cref{tgclass:theorem:approx}, our approximation is suited for large data sets.
To further evaluate the performance on a larger data set, we generated a synthetic data set consisting of 100 graphs generated using the Barabási–Albert preferential attachment method~\cite{barabasi1999emergence} and randomly chosen availability times $t(e)\in [0,1000]$ for each edge $e$. 
Each of the graphs contains $5000$ nodes and $49\,900$ edges with an average maximal degree of $390.3$.
The dissemination process is simulated analogously to the second classification task.
\Cref{table:rnd} shows the running times and accuracies. 
The results suggest very good scalability of our approximation.
The running times are only a fraction of the exact methods, and the accuracy is high.

\noindent\textbf{Q4.}
\begin{figure}[t]
    \centering
    \begin{subfigure}{1\textwidth}
        \centering
        \includegraphics[width=0.95\linewidth]{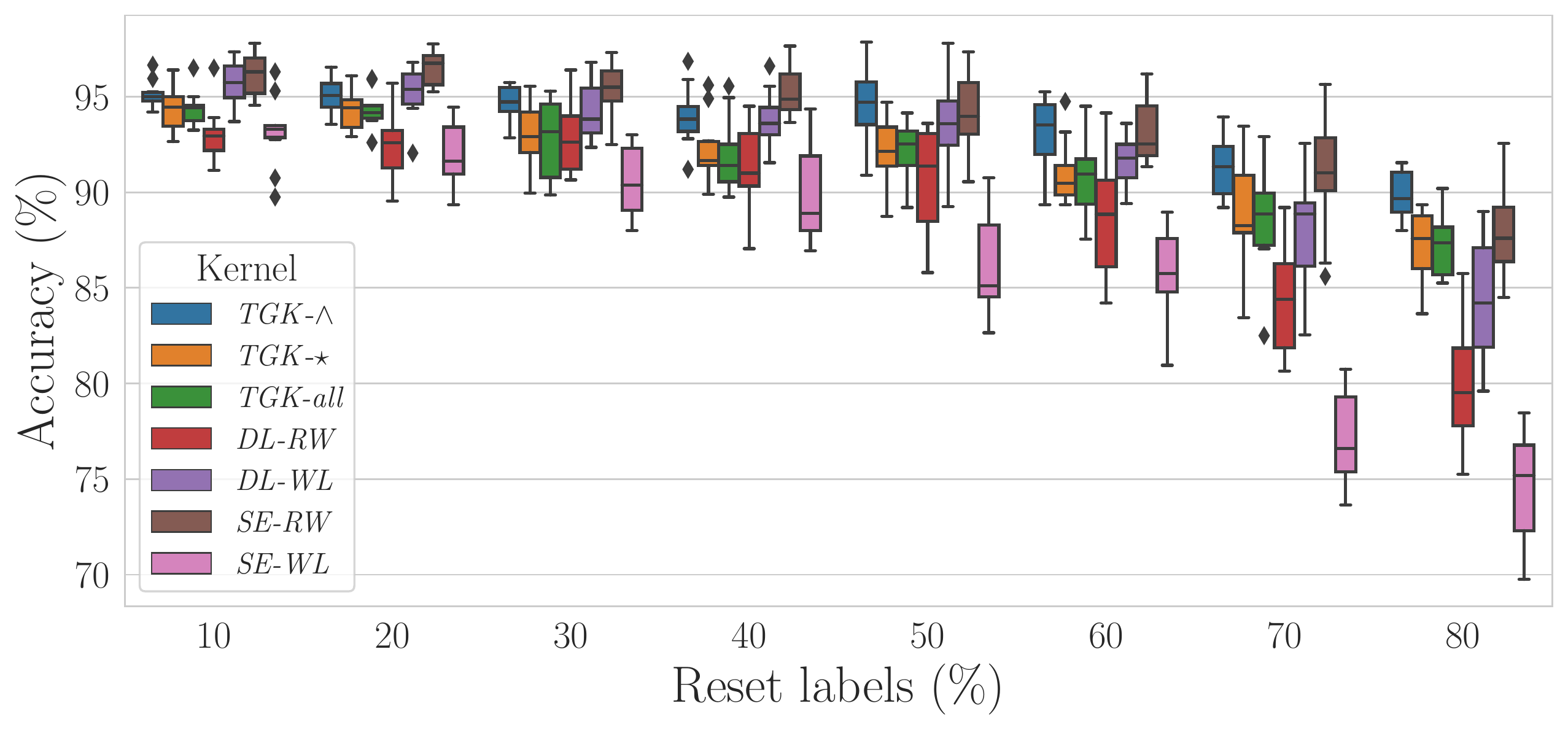}
        \subcaption{Modified first classification task.}
        \label{fig:incomplete_data_task1}        
    \end{subfigure}
    \begin{subfigure}{1\textwidth}
        \centering
        \includegraphics[width=0.95\linewidth]{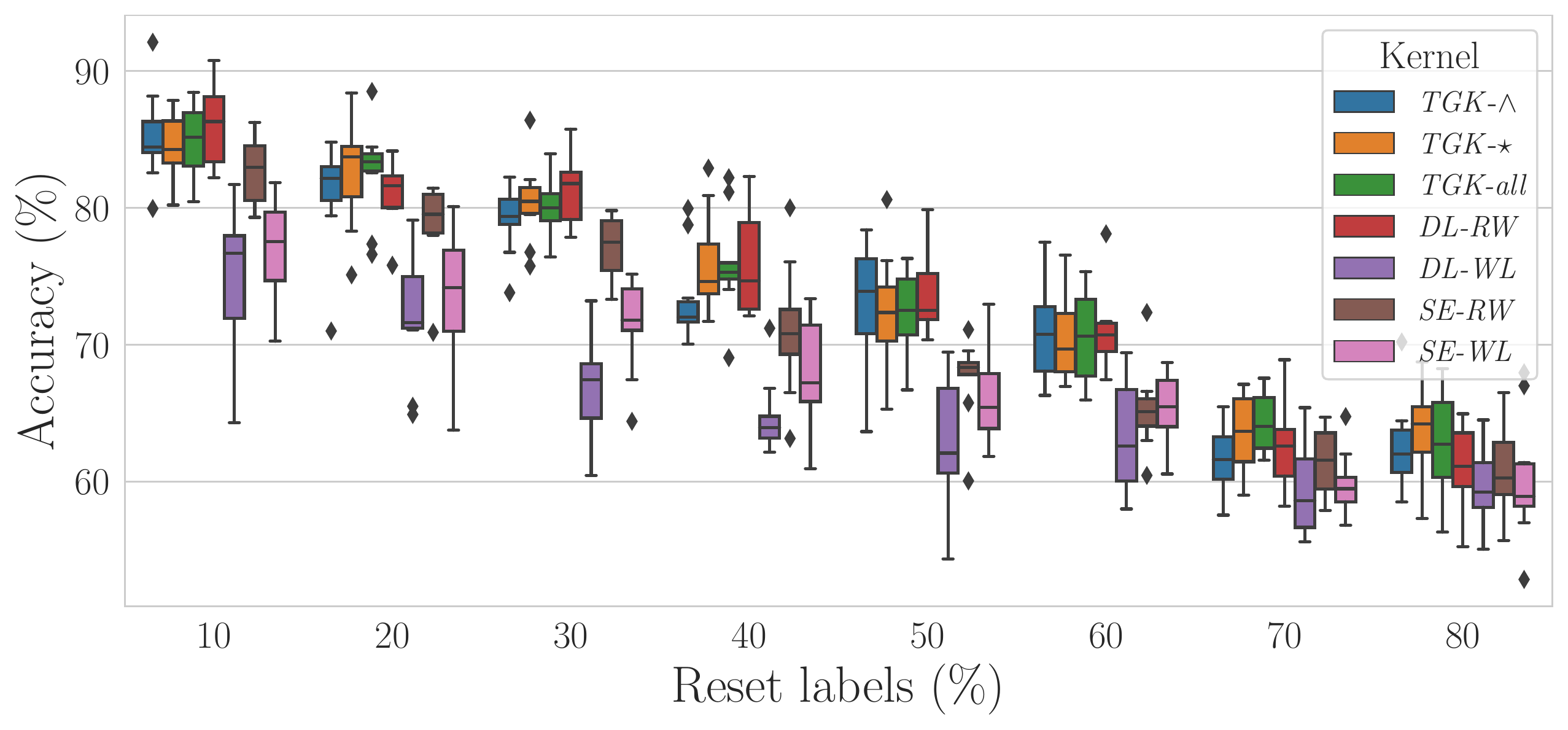}
        \subcaption{Modified second classification task.}
        \label{fig:incomplete_data_task2}
    \end{subfigure}
    \caption{Results for the \textsc{Infectious} data set under incomplete data.}
    \label{fig:incomplete_data}
\end{figure}
In order to evaluate how the accuracy is affected by incomplete knowledge,
we applied our kernel to the third classification task, in which classification tasks one and two are modified by partly reset node labels.
We compare our kernels with the best performing baselines \emph{DL-RW}, \emph{DL-WL}, \emph{SE-RW}, and \emph{SE-WL}.

For the first modified task and $50\%$ to $80\%$ missing data, \tgkwedge has the highest mean accuracy. Our other kernels are on par with the best performing baseline \emph{SE-RW} and are significantly better than the other baselines.
In the case of the second modified task, our kernels beat \emph{SE-RW}, \emph{SE-WL}, and \emph{DL-WL} for all percentages, often with significant gaps in the accuracies, e.g., for the WL-kernels. 
Our kernels have the highest mean accuracy for $20\%$ and from $40\%$ to $80\%$.
The accuracies are on par with the baselines for the remaining percentages. 
In conclusion, our new temporal graphlet kernels often achieve higher mean accuracy than the kernels introduced in~\citeapp{oettershagen2020temporal}, especially in cases where large percentages of information are not available. Hence, our kernels are competitive and often better in the task of classification under missing information.

\section{Conclusion}
We introduced a temporal graphlet kernel for classifying dissemination processes on temporal graphs. Our experimental evaluation showed that our approach beats the state-of-art baselines in most data sets and is on par for the remaining data sets. 
Our approximative kernel has high accuracy while running significantly faster on large data sets. 
The wedge-based kernel has an excellent trade-off between running time and accuracy. 
Finally, our kernels have high accuracy even when information about the dissemination is missing, especially when a majority of the information is unknown. 

\medskip
\noindent\textbf{Acknowledgements}~~
This work is funded by the Deutsche Forschungsgemein\-schaft (DFG, German Research Foundation) under Germany's Excellence Strategy -- EXC-2047/1 -- 390685813.
Nils Kriege has been supported by the Vienna Science and Technology Fund (WWTF) through project VRG19-009.

\appendix

\section{Appendix}

\subsection{Related Work}\label{sec:relatedwork}
Graph kernels have been studied extensively in the past 15 years; see~\citeapp{Kriege2019} for an overview.
Important approaches include random walk and shortest path kernels~\citeapp{Bor+2005,Gaertner2003,KriegeNMKM19}, as well as the Weisfeiler-Leman subtree kernel~\citeapp{Mor+2017,She+2011}. 
Further recent works focus on assignment-based approaches~\citeapp{Kri+2016,Nik+2017}, spectral techniques~\citeapp{Kon+2016}, and graph decomposition~\citeapp{Nik+2018}.
Most kernels are designed for static graphs, with few exceptions considering different aspects of temporal graphs.
Li et al.~\citeapp{Li+2015} present a family of algorithms that efficiently recomputes the random walk kernel when graphs are modified.
Paaßen et al.~\citeapp{Paassen/etal/2017a} use graph kernels for predicting the next graph in a dynamically changing series of graphs.
Similarly, Anil et al.~\citeapp{Anil2014} propose spectral graph kernels to predict the evolution of social networks.

Closely related to our work are \citeapp{oettershagen2020classifying} and \citeapp{oettershagen2020temporal}, which also discuss the classification of dissemination in temporal networks.  To this end, the authors propose transforming the labeled temporal graphs into labeled static graphs.  In \citeapp{oettershagen2020temporal}, they compare three transformations that differ in loss of information and size of the resulting static graph. The authors combine their transformations with the random walk and the Weisfeiler-Leman subtree kernels. The kernel-based approach with the most expressive transformation achieves state-of-the-art classification accuracy but does not scale to large graphs. 
In \citeapp{oettershagen2020classifying}, the transformations have been used with graph neural networks, but the overall performance was worse compared to graph kernels. 
Another closely related work uses dynamic echo state networks~\citeapp{Tortorella2021}.
While the approach does not outperform the accuracy of the transformation-based temporal graph kernels, an advantage is its ability to process graphs in an online setting requiring less space.
Furthermore, various principled extensions of graph neural networks to the temporal domain exist; see this recent survey~\citeapp{KazemiGJKSFP20} and references therein.

Ribeiro et al.~\citeapp{ribeiro2021survey} give an overview of motifs in static networks.
General overviews of temporal graphs can be found in \citeapp{holme2015modern,michail2018elements}.
Various attempts to extend the concept of motifs to evolving graphs have been proposed~\citeapp{DBLP:journals/corr/abs-2005-09721,liu2021temporal}. 
Paranjape et al.~\citeapp{paranjape2017motifs} define temporal motifs as induced subgraphs on sequences of temporal edges.
They introduce an algorithm for counting general temporal graphlets and efficient variants for specific small graphlets.
We generalize these algorithms for counting labeled temporal graphlets.
Mackey et al.~\citeapp{mackey2018chronological} introduced an enumeration algorithm for temporal graphlets.
Several recent works discuss the sampling of temporal motifs~\citeapp{liu2019sampling,liu2021temporal,sarpe2021oden}.

\subsection{Omitted Proofs}
\begin{proof}[Proof of \Cref{theorem:number_of_label_seqs}]
    For each of the $\ell$ temporal edges $(u,v,t)$ in the sequence of temporal edges, we consider the number of different labels for the first node $u$ at time $t$ and the second node $v$ at time $t+1$, leading to $L^2$ possible combinations.
    Having $\ell$ edges leads to $L^{2\ell}$ possible sequences in total. \qed
\end{proof} 
\begin{proof}[Proof of \Cref{lemma:mapping}]
    We use an edge labeling function $\xi$ and for each edge $e=(u,v,t)\in \tge$, we construct a new label $\xi(e)$
    by the concatenation of the time-dependent node labels $l(u,t)\#l(v,t+1)$, where $\#$ is an additional symbol not in $\Sigma$. The edge label sequence is then $l_e(g)=(\xi(e_1),\ldots,\xi(e_\ell))$.
    Hence, there is a one-to-one mapping between $l(g)$ and $l_e(g)$ for all labeled temporal graphlets $g$. \qed
\end{proof}
\begin{proof}[Proof of \Cref{theorem:runtime}]
    If the temporal graph is given in incident list representation, i.e., each vertex has a list of incident temporal edges, and the degree $d(v)$ can be determined in constant time. Then, computing the number of wedges and the probabilities $p_v$ is possible in $\mathcal{O}(|V|)$ time.
    Due to \Cref{theorem:number_of_label_seqs} the initialization takes $\mathcal{O}(|\Sigma|^{2})$ time, and each operation in the for loop is possible in constant time. \qed
\end{proof}
\begin{proof}[Proof of \Cref{tgclass:theorem:approx}]
    First, by an application of the Hoeffding bound together with the union bound, it follows that by setting 
    \begin{equation*}
        s = \frac{\log( 2 \cdot  |\mathbb{G}| \cdot W/\delta)}{ 2\varepsilon^2  },
    \end{equation*}
    it holds that %
    \begin{equation*}
        P\left(| {\phi}_{\text{TG}}(\tg_i)_j - \widetilde{\phi}_{\text{TG}}(\tg_i)_j | \leq\varepsilon\right) \geq 1-\delta,
    \end{equation*} 
    for all $1\leq  j \leq W$, and all temporal graphs $\tg_i$ in $\mathcal{G}$. 
    Let $\tg_1$ and $\tg_2$ in $\mathcal{G}$, then
    \begin{equation*}
        \begin{split}
            \widetilde{k}_{\text{TG}}(\tg_1,\tg_2) &= \left\langle \widetilde{\phi}_{\text{TG}}(\tg_1), \widetilde{\phi}_{\text{TG}}(\tg_2) \right\rangle\\
            &\leq \smashoperator{\sum^{W}_{i = 1}} \mleft({\phi}_{\text{TG}}(\tg_1)_i \cdot{\phi}_{\text{TG}}(\tg_2)_i\mright) + \varepsilon \cdot \smashoperator{\sum^{W}_{i = 1}} \mleft({\phi}_{\text{TG}}(\tg_1)_i + {\phi}_{\text{TG}}(\tg_2)_i \mright) + \smashoperator{\sum^{W}_{i = 1}} \varepsilon^2\\ &\leq {k}_{\text{TG}}(\tg_1,\tg_2) + 2 W  \cdot \varepsilon + W\cdot \varepsilon\,.
        \end{split}
    \end{equation*}
    The last inequality holds because the components of ${\phi}_{\text{TG}}(\cdot)$ are in $[0,1]$. 
    Finally, by setting $\varepsilon = {\lambda}/{W}$ the result follows. \qed
\end{proof}

\subsection{Data Sets}\label{appendix:datasets}
\begin{itemize}%
    \item \textsc{Infectious} and \textsc{Highschool:}
    Contain face-to-face contacts
    between visitors of the exhibition \textit{Infectious: Stay Away}~\cite{Isella2011} and 
    interactions between high school students.
    \item \textsc{Mit:} A temporal graph of contacts between students~\cite{konect:eagle06}. 
    \item \textsc{Facebook} and \textsc{Tumblr:} 
    The first graph is a subset of the activity of a Facebook community over three months~\cite{viswanath2009evolution}. The Tumblr graph contains quoting between Tumblr users~\cite{leskovec2009meme}.  
    \item \textsc{Dblp:} A subset of the Digital Bibliography \& Library Project
    (DBLP) database representing temporal co-author graphs.  
    Nodes represent authors, and the time stamp of an edge is the year of a joint publication. 
\end{itemize}
\begin{table*}[t]\centering	\renewcommand{\arraystretch}{0.55}
    \caption{Statistics and properties of the data sets.} 
    \label{table:sizes1}
    \resizebox{0.8\textwidth}{!}{ 	\renewcommand{\arraystretch}{1}\setlength{\tabcolsep}{5pt}
        \begin{tabular}{lcccccccc}	\toprule
            
            \multirow{3}{*}{\vspace*{4pt}\textbf{Property}}&\multicolumn{6}{c}{\textbf{Data set}}\\\cmidrule{2-7}
            & {\textsc{Mit}}        & \textsc{Highschool}  & \textsc{Infectious}&  \textsc{Tumblr}   & \textsc{Dblp}  & \textsc{Facebook}  \\ \toprule
            
            $\#$ Graphs      &	$97$          &       $180$        &       $200$      &             $373$      &   $755$        &       $995$        \\ %
            Avg. $|V|$     &   $20$           &       $52.3$       &       $50$       &             $53.1$     &     $52.9$     &       $95.7$       \\ 	
            Min $|\tge|$            &   $63$           &       $151$        &       $109$      &             $48$       &     $105$      &       $88$         \\
            Max $|\tge|$            &   $3\,363$       &       $589$        &       $505$      &             $190$      &    $275$       &       $181$        \\ 
            Avg. $|\tge|$  &   $734.6$        &       $272.4$      &       $229.9$    &             $99.9$     &  $160.0$       &       $134.5$      \\ 
            Avg. $\max$ $d(v)$  &   $712.1$     &       $96.8$       &       $46.1$     &             $25.1$     &  $27.1$        &       $21.2$       \\  %
            \bottomrule  \end{tabular}}
\end{table*}
The authors of \cite{oettershagen2020temporal} proposed three classification tasks and provided the labeled data sets.\footnote{\url{http://graphlearning.io/}}
Each classification data set consists of a set of graphs belonging to two classes. 
\Cref{table:sizes1} shows the statistics of the data sets.
The first classification task is the discrimination of observations of a dissemination process and random labeling.
The second classification task aims to discriminate temporal graphs that were subject to two different dissemination processes differing in 
the infection probability.
The third classification task is the classification under incomplete information.
The reasons for missing information about the dissemination process are manifold, e.g., non-symptomatic infections, people hiding their infection for various reasons, or spreading fake news recorded at known spreaders only. 
Based on the \textsc{Infectious} data set and the first two classification tasks, 
for each graph, the labels of $\{10\%,\ldots,80\%\}$ of randomly chosen infected nodes are set back to non-infected.
This process is repeated ten times resulting in 80 data sets for each of the two classification tasks.
\subsection{Experimental Protocol}\label{ep}
We computed the normalized Gram matrices and report the classification accuracies obtained with the $C$-SVM implementation of \text{LIBSVM}~\cite{Cha+2011}, using 10-fold cross-validation. The $C$-parameter was selected from $\{10^{-3}, 10^{-2}, \dotsc, 10^{2},$ $10^{3}\}$ by 10-fold cross-validation on the training folds. 			
We repeated each 10-fold cross-validation ten times with different random folds and report average accuracies and standard deviations.
The time window for the temporal graphlets ($\delta\in\{10^i\mid 1\leq i\leq4 \}$) was selected by fold-wise 10-fold cross-validation. 
Likewise, the number of steps of the random walk kernel ($k\in\{1,\ldots,5\}$) and the number of iterations of the  Weisfeiler-Leman subtree kernel ($h\in\{1,\ldots,5\}$) were selected by fold-wise 10-fold cross-validation. 
All experiments ran on a computer cluster. Each experiment had an exclusive node with an Intel(R) Xeon(R) Gold 6130 CPU @ 2.10GHz and 192 GB of RAM. The time limit was one hour.
Analogous to \cite{oettershagen2020classifying} and to compare the running times, we set the walk length of \emph{DL-RW} to $k=2$ and the number of iterations of \emph{DL-WL} to $h=2$.

\subsection{Running Times for the First Classification Task}\label{appendix:runningtimes}

\Cref{table:runningtimes1} shows the running times for the first classification. 

 \begin{table}
    \centering
    \caption{Running times in ms for the first classification task. Random walk length $k=3$ ($k=2$ for \emph{DL-RW}), number of iterations of WL $h=3$ ($h=2$ for \emph{DL-WL}). OOT---Computation did not finish within the time limit.} 
    \label{table:runningtimes1}
    \resizebox{0.88\textwidth}{!}{ 	\renewcommand{\arraystretch}{0.8}\setlength{\tabcolsep}{5pt}
        \begin{tabular}{cc@{\hspace{5mm}}l@{\hspace{5mm}}ccccccc@{}}	\toprule
            
            & \multirow{3}{*}{\vspace*{4pt}\textbf{Kernel}}&&\multicolumn{6}{c}{\textbf{Data set}}\\\cmidrule{4-9}
            & &  & {\textsc{Mit}}        & \textsc{Highschool}  & \textsc{Infectious}&  \textsc{Tumblr}   & \textsc{Dblp}  & \textsc{Facebook}  \\ \toprule
            & \tgkwedge &    $\delta=10$     & $291$ & $35$ & $40$ & $68$ & $297$ & $443$ \\ 
            & \tgkstar  &    $\delta=10$     & $100$ & $180$ & $191$ & $409$ & $1365$ & $2366$ \\
            & \tgkall   &    $\delta=10$     & $202$ & $793$ & $1407$ & $439$ & $3692$ & $1336$ \\
            \midrule
            & \tgkwedge &   $\delta=10^3$     & $501$ & $59$ & $42$ & $70$ & $290$ & $454$ \\
            & \tgkstar  &   $\delta=10^3$     & $100$ & $180$ & $195$ & $414$ & $1363$ & $2425$ \\
            & \tgkall   &   $\delta=10^3$     & $193$ & $619$ & $1326$ & $374$ & $3540$ & $1234$ \\
             \cmidrule{1-9}
            \multirow{6}{*}{\rotatebox{90}{\small Baselines}}       
            & \emph{RG-RW} & & $793$ & $9272$ & $9153$ & $5040$ & $14428$ & $9700$ \\
            & \emph{RG-WL} & & $18$ & $61$ & $76$ & $109$ & $294$ & $564$ \\
            & \emph{DL-RW} & & OOT & $17871$ & $5683$ & $1552$ & $3532$ & $2322$ \\
            & \emph{DL-WL} & & $56762$ & $1323$ & $574$ & $306$ & $884$ & $787$ \\
            & \emph{SE-RW} & & $3899$ & $1833$ & $2461$ & $973$ & $3370$ & $2810$ \\
            & \emph{SE-WL} & & $338$ & $264$ & $189$ & $209$ & $631$ & $787$ \\
            \bottomrule  \end{tabular}}    
\end{table}

\subsection{Comparison to Graph Neural Networks}\label{appendix:gnn}

We present the classification results using the neural approaches presented in \citeapp{oettershagen2020classifying} and \citeapp{Tortorella2021}.
We compare them to our temporal graphlet kernels.
The GNN model in \citeapp{oettershagen2020classifying} is based on the GIN architecture introduced in~\citeapp{Xu+2018b}. The final GNN layer is fed into a four-layer MLP followed by a softmax function. Alternatively, a Jumping Knowledge (JK) approach~\citeapp{XuXu} to combine the outputs of all layers is used.
The neural networks were trained for $200$ epochs using the Adam optimizer with cross-entropy loss. For each of our three transformations introduced in \citeapp{oettershagen2020temporal}, the following GNNs were trained on the transformed data sets:
\begin{enumerate}
    \item \emph{RG-GIN} and \emph{RG-JK} for the reduced graph, 
    \item \emph{DL-GIN} and \emph{DL-JK} for the directed line graph expansion, and 
    \item \emph{SE-GIN} and \emph{SE-JK} for the static expansion. 
\end{enumerate}
The GNNs in~\citeapp{oettershagen2020classifying} were implemented using the \emph{PyTorch Geometric} library~\citeapp{Fey/Lenssen/2019}.

\emph{DynGESN} is the dynamic echo state network introduced in~\citeapp{Tortorella2021} as an adaption of static graph echo state networks (see~\citeapp{gallicchio2010graph}) for temporal graphs.
The approach is implemented in Matlab\footnote{\url{https://github.com/dtortorella/dyngraphesn}}.

\subsubsection{Results}
\Cref{table:resultsgnn} and \Cref{table:results2gnn} show the results for our kernels and the GNNs. 
Our temporal graphlet kernels beat the neural approaches for all but one data set. In the case of the second classification task, our kernel accuracies are often significantly higher compared to the GNN accuracies.

\begin{table}[tbh]\centering
    \caption{Classification accuracy in percent and standard deviation for the first and second classification task. For each data set, we highlight the highest accuracy in bold. OOM---Out of memory.}
    \begin{subtable}{1\textwidth}
        \caption{Classification accuracy for the first classification task.} %
        \label{table:resultsgnn}
        \resizebox{1\textwidth}{!}{ 	\renewcommand{\arraystretch}{0.8}\setlength{\tabcolsep}{5pt}
            \begin{tabular}{cl@{\hspace{5mm}}ccccccc}	\toprule
                & \multirow{3}{*}{\vspace*{4pt}\textbf{Kernel}}&\multicolumn{6}{c}{\textbf{Data set}}\\\cmidrule{3-8}
                &  & {\textsc{Mit}}         &  {\textsc{Highschool}}      &     {\textsc{Infectious}}       & {\textsc{Tumblr}}       & {\textsc{Dblp}} &  {\textsc{Facebook}} \\ \toprule
                & \tgkwedge    & $\textbf{93.45}$ $\sz\pm 1.8$ & $\textbf{98.00}$ $\sz\pm 0.9$ & $\textbf{98.30}$ $\sz\pm 0.4$ & $93.19$ $\sz\pm 0.7$ & $\textbf{98.47}$ $\sz\pm 0.1$ &  $95.21$ $\sz\pm 0.2$\\
                & \tgkstar     & $87.02$ $\sz\pm 1.6$ & $96.50$ $\sz\pm 0.9$ & ${98.05}$ $\sz\pm 0.4$ & $93.19$ $\sz\pm 0.8$ & $98.05$ $\sz\pm 0.2$ & $\textbf{95.32}$ $\sz\pm 0.2$\\
                & \tgkall      & $87.46$ $\sz\pm 0.9$ & $97.17$ $\sz\pm 0.8$ & $97.90$ $\sz\pm 0.4$ & $\textbf{93.57}$ $\sz\pm 0.8$ & $98.12$ $\sz\pm 0.1$ & $94.69$ $\sz\pm 0.3$\\ 
                \midrule\multirow{7}{*}{\rotatebox{90}{\small Baselines}}
                & \emph{RG-GIN} & $50.65$ $\sz\pm 4.2~$& $51.11$ $\sz\pm 2.5~$ & $58.20$ $\sz\pm 4.0~$ & $72.63$ $\sz\pm 1.8~$ & $86.36$ $\sz\pm 0.9~$ & $89.54$ $\sz\pm 0.8~$\\
                & \emph{RG-JK}  & $50.74$ $\sz\pm 3.1~$& $50.83$ $\sz\pm 4.9~$ & $47.85$ $\sz\pm 2.7~$ & $69.14$ $\sz\pm 3.6~$ & $86.43$ $\sz\pm 0.7~$ & $87.27$ $\sz\pm 0.6~$\\
                & \emph{DL-GIN}& OOM                 & $88.67$ $\sz\pm 2.1~$ & $92.85$ $\sz\pm 1.7~$ & $90.39$ $\sz\pm 1.4~$ & $97.72$ $\sz\pm 0.4~$ & $94.29$ $\sz\pm 0.2~$ \\
                & \emph{DL-JK} & OOM                  & $86.22$ $\sz\pm 2.6~$ & $91.55$ $\sz\pm 2.3~$ & $89.30$ $\sz\pm 1.5~$ & $97.57$ $\sz\pm 0.3~$ & $93.05$ $\sz\pm 0.6~$ \\
                & \emph{SE-GIN} & $75.98$ $\sz\pm 3.7~$& $92.28$ $\sz\pm 1.2~$ & $93.10$ $\sz\pm 1.9~$ & $92.78$ $\sz\pm 1.1~$ & $97.87$ $\sz\pm 0.3~$ & $94.72$ $\sz\pm 0.5~$ \\
                & \emph{SE-JK}  & $75.37$ $\sz\pm 3.6~$& $92.33$ $\sz\pm 2.7~$ & $93.50$ $\sz\pm 1.9~$ & $92.30$ $\sz\pm 0.9~$ & $97.14$ $\sz\pm 0.9~$ & $95.02$ $\sz\pm 0.6~$ \\
                & \emph{DynGESN} & $69.70$ $\sz\pm 9.2~$ & $94.40$ $\sz\pm 5.3~$ & $94.70$ $\sz\pm 5.3~$ & $93.30$ $\sz\pm 3.9~$ & $97.70 $ $\sz\pm 1.8~$ & $93.00 $ $\sz\pm 2.5~$ \\	
                \bottomrule  \end{tabular}}
    \end{subtable}
    \begin{subtable}{1\textwidth}
        \caption{Classification accuracy for the second classification task.  
        } %
    \label{table:results2gnn}
    \resizebox{1\textwidth}{!}{ 	\renewcommand{\arraystretch}{0.8}\setlength{\tabcolsep}{5pt}
        \begin{tabular}{clccccccc}	\toprule
            & \multirow{3}{*}{\vspace*{4pt}\textbf{Kernel}}&\multicolumn{6}{c}{\textbf{Data set}}\\\cmidrule{3-8}
            &  & {\textsc{Mit}}         &  {\textsc{Highschool}}      &     {\textsc{Infectious}}       & {\textsc{Tumblr}}       & {\textsc{Dblp}} &  {\textsc{Facebook}} \\ \toprule
            & \tgkwedge    & $68.52$ $\sz\pm 3.5$ & $93.83$ $\sz\pm 0.8$ & $89.65$ $\sz\pm 0.8$ & $\textbf{79.06}$ $\sz\pm 0.7$ & $83.76$ $\sz\pm 0.5$ & $76.63$ $\sz\pm 0.3$\\
            & \tgkstar     & $77.03$ $\sz\pm 3.7$ & $\textbf{95.33}$ $\sz\pm 0.8$ & $\textbf{90.55}$ $\sz\pm 1.6$ & ${78.13}$ $\sz\pm 1.1$ & $85.42$ $\sz\pm 0.8$ & ${82.12}$ $\sz\pm 0.3$ \\
            & \tgkall      & $\textbf{78.43}$ $\sz\pm 4.2$ & $94.22$ $\sz\pm 1.3$ & $90.40$ $\sz\pm 0.9$ & $76.57$ $\sz\pm 1.0$ & $\textbf{85.84}$ $\sz\pm 0.6$ & $81.13$ $\sz\pm 0.6$ \\\midrule
            \multirow{7}{*}{\rotatebox{90}{\small Baselines}}
            & \emph{RG-GIN} & $53.80$ $\sz\pm 16.0$& $53.61$ $\sz\pm 3.5~$ & $51.80$ $\sz\pm 4.2~$ & $64.70$ $\sz\pm 2.4~$ & $60.24$ $\sz\pm 1.8~$ & $67.75$ $\sz\pm 1.3~$ \\
            & \emph{RG-JK}  & $51.80$ $\sz\pm 9.7~$& $54.61$ $\sz\pm 2.9~$ & $52.60$ $\sz\pm 3.2~$ & $65.50$ $\sz\pm 2.6~$ & $61.00$ $\sz\pm 1.1~$ & $67.75$ $\sz\pm 1.0~$ \\
            & \emph{DL-GIN} & OOM                  & $89.11$ $\sz\pm 2.0~$ & $80.60$ $\sz\pm 2.2~$ & $75.45$ $\sz\pm 2.4~$ & $80.05$ $\sz\pm 1.1~$ & $\textbf{83.16}$ $\sz\pm 0.9~$ \\
            & \emph{DL-JK}  & OOM                  & $85.00$ $\sz\pm 2.8~$ & $75.70$ $\sz\pm 3.5~$ & $73.10$ $\sz\pm 1.8~$ & $79.98$ $\sz\pm 1.3~$ & $82.26$ $\sz\pm 1.1~$ \\
            & \emph{SE-GIN} & $51.40$ $\sz\pm11.1$ & $85.88$ $\sz\pm 2.1~$ & $75.05$ $\sz\pm 3.4~$ & $73.23$ $\sz\pm 1.7~$ & $80.72$ $\sz\pm 1.1~$ & $82.21$ $\sz\pm 0.9~$ \\
            & \emph{SE-JK} & $51.40$ $\sz\pm 10.9$ & $82.44$ $\sz\pm 2.0~$ & $74.25$ $\sz\pm 2.4~$ & $74.55$ $\sz\pm 1.4~$ & $81.18$ $\sz\pm 1.0~$ & $80.46$ $\sz\pm 0.9$ \\
            & \emph{DynGESN} & $63.30 $ $\sz\pm 11.0$ & $92.80$ $\sz\pm 5.2~$ & $80.60$ $\sz\pm 9.1~$ & $76.80$ $\sz\pm 6.2~$ & $74.30$ $\sz\pm 4.7~$ & $76.10$ $\sz\pm 3.9~$ \\	
            \bottomrule  \end{tabular}}
\end{subtable}
\end{table}


\begin{thebibliography}{10}
    \providecommand{\url}[1]{\texttt{#1}}
    \providecommand{\urlprefix}{URL }
    \providecommand{\doi}[1]{https://doi.org/#1}
    
    \bibitem{barabasi1999emergence}
    Barab{\'a}si, A.L., Albert, R.: Emergence of scaling in random networks.
    science  \textbf{286}(5439),  509--512 (1999)
    
    \bibitem{braha2009time}
    Braha, D., Bar-Yam, Y.: Time-Dependent Complex Networks: Dynamic Centrality,
    Dynamic Motifs, and Cycles of Social Interactions, pp. 39--50. Springer
    Berlin Heidelberg, Berlin, Heidelberg (2009)
    
    \bibitem{brauer2008compartmental}
    Brauer, F.: Compartmental models in epidemiology. In: Mathematical
    epidemiology, pp. 19--79. Springer (2008)
    
    \bibitem{brouwer2018epidemiology}
    Brouwer, A.F., Eisenberg, J.N., Pomeroy, C.D., Shulman, L.M., Hindiyeh, M.,
    Manor, Y., Grotto, I., Koopman, J.S., Eisenberg, M.C.: Epidemiology of the
    silent polio outbreak in rahat, israel, based on modeling of environmental
    surveillance data. Proc. of the National Academy of Sciences
    \textbf{115}(45),  E10625--E10633 (2018)
    
    \bibitem{Cha+2011}
    Chang, C.C., Lin, C.J.: {LIBSVM}: {A} library for support vector machines. ACM
    Transactions on Intelligent Systems and Technology  \textbf{2},  27:1--27:27
    (2011)
    
    \bibitem{konect:eagle06}
    Eagle, N., Pentland, A.S.: Reality mining: Sensing complex social systems.
    Personal and Ubiquitous Computing  \textbf{10}(4),  255--268 (2006)
    
    \bibitem{holme2015modern}
    Holme, P.: Modern temporal network theory: A colloquium. The European Physical
    Journal B  \textbf{88}(9), ~234 (2015)
    
    \bibitem{Isella2011}
    Isella, L., Stehlé, J., Barrat, A., Cattuto, C., Pinton, J.F., Van~den Broeck,
    W.: What's in a crowd? {A}nalysis of face-to-face behavioral networks.
    Journal of Theoretical Biology  \textbf{271}(1),  166--180 (2011)
    
    \bibitem{kaslow2014viral}
    Kaslow, R.A., Stanberry, L.R., Le~Duc, J.W.: Viral infections of humans:
    epidemiology and control. Springer (2014)
    
    \bibitem{Kriege2019}
    Kriege, N.M., Johansson, F.D., Morris, C.: A survey on graph kernels. Applied
    Network Science  \textbf{5}(1),  1--42 (2020)
    
    \bibitem{leskovec2009meme}
    Leskovec, J., Backstrom, L., Kleinberg, J.: Meme-tracking and the dynamics of
    the news cycle. In: Proceedings of the 15th ACM SIGKDD International
    Conference on Knowledge Discovery and Data Mining. pp. 497--506 (2009)
    
    \bibitem{liu2019sampling}
    Liu, P., Benson, A.R., Charikar, M.: Sampling methods for counting temporal
    motifs. In: Proceedings of the twelfth ACM international conference on web
    search and data mining. pp. 294--302 (2019)
    
    \bibitem{liu2021temporal}
    Liu, P., Guarrasi, V., Sariyuce, A.E.: Temporal network motifs: Models,
    limitations, evaluation. IEEE Transactions on Knowledge and Data Engineering
    (2021)
    
    \bibitem{mackey2018chronological}
    Mackey, P., Porterfield, K., Fitzhenry, E., Choudhury, S., Chin, G.: A
    chronological edge-driven approach to temporal subgraph isomorphism. In: 2018
    IEEE international conference on big data (big data). pp. 3972--3979. IEEE
    (2018)
    
    \bibitem{masuda2019detecting}
    Masuda, N., Holme, P.: Detecting sequences of system states in temporal
    networks. Scientific Reports  \textbf{9}(1),  1--11 (2019)
    
    \bibitem{murayama2021modeling}
    Murayama, T., Wakamiya, S., Aramaki, E., Kobayashi, R.: Modeling the spread of
    fake news on twitter. Plos one  \textbf{16}(4),  e0250419 (2021)
    
    \bibitem{oettershagen2020classifying}
    Oettershagen, L., Kriege, N.M., Morris, C., Mutzel, P.: Classifying
    dissemination processes in temporal graphs. Big Data  \textbf{8}(5),
    363--378 (2020)
    
    \bibitem{oettershagen2020temporal}
    Oettershagen, L., Kriege, N.M., Morris, C., Mutzel, P.: Temporal graph kernels
    for classifying dissemination processes. In: Proceedings of the 2020 SIAM
    International Conference on Data Mining. pp. 496--504. SIAM (2020)
    
    \bibitem{palladino2020excess}
    Palladino, R., Bollon, J., Ragazzoni, L., Barone-Adesi, F.: Excess deaths and
    hospital admissions for covid-19 due to a late implementation of the lockdown
    in italy. Intl. Journal of Environmental Research and Public Health
    \textbf{17}(16), ~5644 (2020)
    
    \bibitem{paranjape2017motifs}
    Paranjape, A., Benson, A.R., Leskovec, J.: Motifs in temporal networks. In:
    Proc. of the ACM Intl. Conf. on Web Search and Data Mining. pp. 601--610
    (2017)
    
    \bibitem{sarpe2021oden}
    Sarpe, I., Vandin, F.: oden: Simultaneous approximation of multiple motif
    counts in large temporal networks. In: Proceedings of the 30th ACM
    International Conference on Information \& Knowledge Management. pp.
    1568--1577 (2021)
    
    \bibitem{seshadhri2014wedge}
    Seshadhri, C., Pinar, A., Kolda, T.G.: Wedge sampling for computing clustering
    coefficients and triangle counts on large graphs. Statistical Analysis and
    Data Mining: The ASA Data Science Journal  \textbf{7}(4),  294--307 (2014)
    
    \bibitem{shervashidze2009efficient}
    Shervashidze, N., Vishwanathan, S., Petri, T., Mehlhorn, K., Borgwardt, K.:
    Efficient graphlet kernels for large graph comparison. In: Artificial
    intelligence and statistics. pp. 488--495. PMLR (2009)
    
    \bibitem{shu2017fake}
    Shu, K., Sliva, A., Wang, S., Tang, J., Liu, H.: Fake news detection on social
    media: A data mining perspective. ACM SIGKDD explorations newsl.
    \textbf{19}(1),  22--36 (2017)
    
    \bibitem{Tortorella2021}
    Tortorella, D., Micheli, A.: Dynamic graph echo state networks. In: Proceedings
    of the 29th European Symposium on Artificial Neural Networks, Computational
    Intelligence and Machine Learning (ESANN). pp. 99--104 (2021)
    
    \bibitem{uvzupyte2020test}
    U{\v{z}}upyt{\.e}, R., Wit, E.C.: Test for triadic closure and triadic
    protection in temporal relational event data. Social Network Analysis and
    Mining  \textbf{10}(1),  1--12 (2020)
    
    \bibitem{viswanath2009evolution}
    Viswanath, B., Mislove, A., Cha, M., Gummadi, K.P.: On the evolution of user
    interaction in facebook. In: Proc. ACM Works. on Onl. Soc. Net. pp. 37--42
    (2009)
    
    \bibitem{vosoughi2018spread}
    Vosoughi, S., Roy, D., Aral, S.: The spread of true and false news online.
    Science  \textbf{359}(6380),  1146--1151 (2018)
    
    \bibitem{wale2008comparison}
    Wale, N., Watson, I.A., Karypis, G.: Comparison of descriptor spaces for
    chemical compound retrieval and classification. Knowl. and Inf. Sys.
    \textbf{14}(3),  347--375 (2008)
    
    \bibitem{wu2019comprehensive}
    Wu, Z., Pan, S., Chen, F., Long, G., Zhang, C., Yu, P.S.: A comprehensive
    survey on graph neural networks. {IEEE} Trans. Neural Net. Learn. Syst.
    \textbf{32}(1),  4--24 (2021)
    
\end{thebibliography}

\begin{thebibliography}{10}
    \providecommand{\url}[1]{\texttt{#1}}
    \providecommand{\urlprefix}{URL }
    \providecommand{\doi}[1]{https://doi.org/#1}
    
    \bibitem{Anil2014}
    Anil, A., Sett, N., Singh, S.R.: Modeling evolution of a social network using
    temporalgraph kernels. In: Proceedings of the 37th International ACM SIGIR
    Conference on Research \& Development in Information Retrieval. pp.
    1051--1054 (2014)
    
    \bibitem{Bor+2005}
    Borgwardt, K.M., Kriegel, H.P.: Shortest-path kernels on graphs. In: Fifth IEEE
    International Conference on Data Mining. pp. 8--pp. IEEE (2005)
    
    \bibitem{Fey/Lenssen/2019}
    Fey, M., Lenssen, J.E.: Fast graph representation learning with {PyTorch
        Geometric}. In: ICLR Workshop on Representation Learning on Graphs and
    Manifolds (2019)
    
    \bibitem{gallicchio2010graph}
    Gallicchio, C., Micheli, A.: Graph echo state networks. In: The 2010
    international joint conference on neural networks (IJCNN). pp.~1--8. IEEE
    (2010)
    
    \bibitem{Gaertner2003}
    G{\"a}rtner, T., Flach, P., Wrobel, S.: On graph kernels: Hardness results and
    efficient alternatives. In: Learning Theory and Kernel Mach., pp. 129--143.
    Springer (2003)
    
    \bibitem{holme2015modern}
    Holme, P.: Modern temporal network theory: A colloquium. The European Physical
    Journal B  \textbf{88}(9), ~234 (2015)
    
    \bibitem{DBLP:journals/corr/abs-2005-09721}
    Jazayeri, A., Yang, C.C.: Motif discovery algorithms in static and temporal
    networks: {A} survey. CoRR  \textbf{abs/2005.09721} (2020)
    
    \bibitem{KazemiGJKSFP20}
    Kazemi, S.M., Goel, R., Jain, K., Kobyzev, I., Sethi, A., Forsyth, P., Poupart,
    P.: Representation learning for dynamic graphs: {A} survey. J. Mach. Learn.
    Res.  \textbf{21},  70:1--70:73 (2020)
    
    \bibitem{Kon+2016}
    Kondor, R., Pan, H.: The multiscale {L}aplacian graph kernel. In: Advances in
    Neural Information Processing Systems. pp. 2990--2998 (2016)
    
    \bibitem{Kri+2016}
    Kriege, N.M., Giscard, P.L., Wilson, R.: On valid optimal assignment kernels
    and applications to graph classification. In: Advances in Neural Information
    Processing Systems. pp. 1623--1631 (2016)
    
    \bibitem{Kriege2019}
    Kriege, N.M., Johansson, F.D., Morris, C.: A survey on graph kernels. Applied
    Network Science  \textbf{5}(1),  1--42 (2020)
    
    \bibitem{KriegeNMKM19}
    Kriege, N.M., Neumann, M., Morris, C., Kersting, K., Mutzel, P.: A unifying
    view of explicit and implicit feature maps of graph kernels. Data Mining and
    Knowledge Discovery  \textbf{33}(6),  1505--1547 (2019)
    
    \bibitem{Li+2015}
    Li, L., Tong, H., Xiao, Y., Fan, W.: Cheetah: Fast graph kernel tracking on
    dynamic graphs. In: Proceedings of the 2015 SIAM International Conference on
    Data Mining. pp. 280--288. SIAM (2015)
    
    \bibitem{liu2019sampling}
    Liu, P., Benson, A.R., Charikar, M.: Sampling methods for counting temporal
    motifs. In: Proceedings of the twelfth ACM international conference on web
    search and data mining. pp. 294--302 (2019)
    
    \bibitem{liu2021temporal}
    Liu, P., Guarrasi, V., Sariyuce, A.E.: Temporal network motifs: Models,
    limitations, evaluation. IEEE Transactions on Knowledge and Data Engineering
    (2021)
    
    \bibitem{mackey2018chronological}
    Mackey, P., Porterfield, K., Fitzhenry, E., Choudhury, S., Chin, G.: A
    chronological edge-driven approach to temporal subgraph isomorphism. In: 2018
    IEEE international conference on big data (big data). pp. 3972--3979. IEEE
    (2018)
    
    \bibitem{michail2018elements}
    Michail, O., Spirakis, P.G.: Elements of the theory of dynamic networks.
    Communications of the ACM  \textbf{61}(2),  72--72 (2018)
    
    \bibitem{Mor+2017}
    Morris, C., Kersting, K., Mutzel, P.: Glocalized {W}eisfeiler-{L}ehman graph
    kernels: Global-local feature maps of graphs. In: 2017 IEEE International
    Conference on Data Mining. pp. 327--336. IEEE (2017)
    
    \bibitem{Nik+2018}
    Nikolentzos, G., Meladianos, P., Limnios, S., Vazirgiannis, M.: A degeneracy
    framework for graph similarity. In: Proceedings of the Twenty-Seventh
    International Joint Conference on Artificial Intelligence. pp. 2595--2601
    (2018)
    
    \bibitem{Nik+2017}
    Nikolentzos, G., Meladianos, P., Vazirgiannis, M.: Matching node embeddings for
    graph similarity. In: Thirty-First AAAI Conference on Artificial Intelligence
    (2017)
    
    \bibitem{oettershagen2020classifying}
    Oettershagen, L., Kriege, N.M., Morris, C., Mutzel, P.: Classifying
    dissemination processes in temporal graphs. Big Data  \textbf{8}(5),
    363--378 (2020)
    
    \bibitem{oettershagen2020temporal}
    Oettershagen, L., Kriege, N.M., Morris, C., Mutzel, P.: Temporal graph kernels
    for classifying dissemination processes. In: Proceedings of the 2020 SIAM
    International Conference on Data Mining. pp. 496--504. SIAM (2020)
    
    \bibitem{Paassen/etal/2017a}
    Paa{\ss}en, B., G{\"o}pfert, C., Hammer, B.: Time series prediction for graphs
    in kernel and dissimilarity spaces. Neural Processing Letters
    \textbf{48}(2),  669--689 (2018)
    
    \bibitem{paranjape2017motifs}
    Paranjape, A., Benson, A.R., Leskovec, J.: Motifs in temporal networks. In:
    Proc. of the ACM Intl. Conf. on Web Search and Data Mining. pp. 601--610
    (2017)
    
    \bibitem{ribeiro2021survey}
    Ribeiro, P., Paredes, P., Silva, M.E., Aparicio, D., Silva, F.: A survey on
    subgraph counting: Concepts, algorithms, and applications to network motifs
    and graphlets. ACM Computing Surveys (CSUR)  \textbf{54}(2),  1--36 (2021)
    
    \bibitem{sarpe2021oden}
    Sarpe, I., Vandin, F.: oden: Simultaneous approximation of multiple motif
    counts in large temporal networks. In: Proceedings of the 30th ACM
    International Conference on Information \& Knowledge Management. pp.
    1568--1577 (2021)
    
    \bibitem{She+2011}
    Shervashidze, N., Schweitzer, P., Van~Leeuwen, E.J., Mehlhorn, K., Borgwardt,
    K.M.: Weisfeiler-{L}ehman graph kernels. Journal of Machine Learning Research
    \textbf{12}(77),  2539--2561 (2011)
    
    \bibitem{Tortorella2021}
    Tortorella, D., Micheli, A.: Dynamic graph echo state networks. In: Proceedings
    of the 29th European Symposium on Artificial Neural Networks, Computational
    Intelligence and Machine Learning (ESANN). pp. 99--104 (2021)
    
    \bibitem{Xu+2018b}
    Xu, K., Hu, W., Leskovec, J., Jegelka, S.: How powerful are graph neural
    networks? arXiv preprint arXiv:1810.00826  (2018)
    
    \bibitem{XuXu}
    Xu, K., Li, C., Tian, Y., Sonobe, T., Kawarabayashi, K., Jegelka, S.:
    Representation learning on graphs with jumping knowledge networks. In: 35th
    International Conference on Machine Learning. pp. 5449--5458 (2018)
    
\end{thebibliography}
\end{document}